\documentclass[]{aa}  

\usepackage{tabularx}
\usepackage[figuresright]{rotating}
\usepackage{txfonts}
\usepackage{graphicx}
\usepackage{nicefrac,xfrac}
\usepackage{upgreek}
\graphicspath{{images}{figs}}
\usepackage[breaklinks=true]{hyperref}
\usepackage{mathrsfs}
\usepackage{tikz}
\usepackage{nicefrac}
\usepackage{orcidlink}

\begin{document} 

   \title{Discovery of large-scale radio emission enveloping the  mini-halo in the most X-ray luminous galaxy cluster RX~J1347.5-1145}
	\author{D. N. Hoang\inst{\ref{Hamburg},\ref{ASIAA},\ref{TLS}}\orcidlink{0000-0002-8286-646X}
	\and
	M. Br\"uggen \inst{\ref{Hamburg}}\orcidlink{0000-0002-3369-7735}
    \and
    A. Bonafede \inst{\ref{Bologna},\ref{INAF}}\orcidlink{0000-0002-5068-4581} 
    \and
    P. M. Koch\inst{\ref{ASIAA}}\orcidlink{0000-0003-2777-5861}
    \and
    G. Brunetti\inst{\ref{INAF}}\orcidlink{0000-0003-4195-8613}
    \and
    E. Bulbul\inst{\ref{MPE}}\orcidlink{0000-0002-7619-5399}
    \and
    G. Di Gennaro\inst{\ref{Hamburg},\ref{INAF}}\orcidlink{0000-0002-8648-8507}
    \and
    A. Liu\inst{\ref{MPE},\ref{BNU}}\orcidlink{0000-0003-3501-0359}
    \and
    C. J. Riseley\inst{\ref{Bologna},\ref{INAF}}\orcidlink{0000-0002-3369-1085}   
    \and
    H. J. A. R\"ottgering \inst{\ref{Leiden}}\orcidlink{0000-0001-8887-2257}
    \and
    R. J. van Weeren\inst{\ref{Leiden}}\orcidlink{0000-0002-0587-1660}
    }
	\authorrunning{D. N. Hoang et al.}
	\titlerunning{Large-scale radio emission enveloping the mini-halo of RX~J1347.5-1145}
	%
	\institute{Hamburger Sternwarte, Universit\"at Hamburg, Gojenbergsweg 112, 21029 Hamburg, Germany\label{Hamburg}
    \and
    Academia Sinica Institute of Astronomy and Astrophysics No. 1, Section 4, Roosevelt Road, Taipei 10617, Taiwan \label{ASIAA}
    \and
    Th\"uringer Landessternwarte, Sternwarte 5, 07778 Tautenburg, Germany \label{TLS}
    \and
    Dipartimento di Fisica e Astronomia, Universit\"a di Bologna, via Gobetti 93/2, 40122 Bologna, Italy \label{Bologna}
    \and
    INAF - Istituto di Radioastronomia di Bologna, Via Gobetti 101, I-40129 Bologna, Italy \label{INAF}
    \and
    Max Planck Institute for Extraterrestrial Physics, Giessenbachstrasse 1, 85748 Garching, Germany \label{MPE} 
    \and
    Institute for Frontiers in Astronomy and Astrophysics, Beijing Normal University, Beijing 102206, China \label{BNU}
    \and
	Leiden Observatory, Leiden University, PO Box 9513, NL-2300 RA Leiden, The Netherlands \label{Leiden}
	}
   \date{Received: 2024; accepted: 2024}

  \abstract
  {Diffuse radio sources, known as mini-halos and halos, are detected at the centres of galaxy clusters. These centralized diffuse sources are typically observed individually, with both appearing together only in rare cases. The origin of the diffuse radio sources in such systems remains unclear.}
  {We investigate the formation of large-scale radio emission in the most X-ray luminous, massive galaxy cluster RXJ~1347.5-1145 which is known to host a mini-halo at its centre and possibly additional more extended emission.}
  {We conduct deep multi-frequency observations of the galaxy cluster using the MeerKAT at 1.28 GHz and the uGMRT (upgraded Giant Metrewave Radio Telescope) at 1.26 GHz and 700 MHz. We characterize the brightness and spectral properties of the central diffuse sources and combine our radio observations with \textit{Chandra} X-ray data to explore the correlation between the cluster's non-thermal and thermal emissions.}
  {We confirm the presence of the diffuse emission and find that it extends up to 1~Mpc in size. Our multi-wavelength data reveal that the central diffuse emission consists of two distinct components: a mini-halo located in the cluster core and a larger radio halo extending around it. The correlation between radio and X-ray surface brightness in both sources indicates a strong connection between the non-thermal and thermal properties of the ICM. The differing slopes in the $I_R-I_X$ and $\alpha-I_X$ relations suggest that distinct mechanisms are responsible for the formation of the mini-halo and halo.  The properties of the halo align with the turbulent model, while both turbulent and hadronic processes may contribute to the formation of the mini-halo.}
  {}


   \maketitle
%

\section{Introduction}
\label{sec:intro}

Galaxy clusters are gravitationally bound systems, containing hundreds to thousands of individual galaxies \citep[e.g.][]{Abell1989,Wen2012}. In the context of hierarchical formation, massive ($M\approx10^{14}-10^{15}$~$M_{\odot}$) clusters form through a series of energetic mergers of smaller clusters that releases up to $\sim$$10^{64}$~erg during a timescale of a few Giga-years \citep[e.g.][]{Hoeft2008}. The cluster-cluster collisions generate turbulence and shocks in the intra-cluster medium (ICM) that are observable with X-ray and radio telescopes \citep[e.g.][]{Brunetti2014, VanWeeren2019a}.

Most of the gravitational energy released during the cluster formation is converted into the thermal motions of particles, which emit Bremsstrahlung X-ray emission in the ICM. Consequently, the morphology of the diffuse X-ray emission serves as a reliable tracer for cluster dynamics \citep[e.g.][]{Buote1995,Poole2006,Santos2008}.  Simultaneously, a small portion of this gravitational energy is channelled into non-thermal components through the acceleration of cosmic ray electrons (CRes) and the amplification of magnetic fields, resulting in diffuse synchrotron emission in the ICM \citep[e.g.][]{Ensslin1998,VanWeeren2010a,Cassano2013a}.

Central diffuse radio sources have been increasingly detected in galaxy clusters, thanks to advancements of low-frequency radio telescopes. These sources are classified as mega-halos, halos, or mini-halos, depending on their observed properties such as their sizes and synchrotron emissivity. These diffuse synchrotron sources exhibit steep ($\alpha\lesssim-1$) spectra\footnote{$S\propto\nu^{\alpha}$ convention is used throughout this paper.} and are unpolarised \citep[up to a few percent; e.g.,][]{Feretti2001}. Radio halos typically span about $\sim$1~Mpc in size (or largest linear size - LLS), have radio surface brightness (SB) profiles following an exponential function, and are typically detected in disturbed galaxy clusters \citep[for a  review, see, e.g., ][]{VanWeeren2019a}. A few halos with exceptionally larger sizes have also been observed in massive merging clusters including Abell 2142, Abell 2163, and 1E0657-56 with a LLS of 2.4~Mpc, 2.9~Mpc and 2.1~Mpc, respectively \citep{Bruno2023a,Liang2000,Feretti2001}. The unusually-large sizes of these radio halos might be due to the high-end masses of the dynamically-disturbed host clusters \citep[i.e., mass of $(8.77\pm0.19)\times10^{14}M_{\rm \odot}$ for Abell~2142, $(1.61\pm0.03)\times10^{15}M_{\rm \odot}$ for Abell~2163 and $(1.31\pm0.03)\times10^{15}M_{\rm \odot}$ for 1E0657-56;][]{PlanckCollaborationXXIV2015} that release more energy during the cluster-cluster merger. Recently, a few clusters have been identified to host centralised diffuse emission with a LLS of larger than 2~Mpc using the LOw Frequency Array (LOFAR) at 144~MHz \citep{Cuciti2022a}. These diffuse sources are found to surround the existing radio halos and have a lower emissivity than the halos by a factor of $\sim20$; hence they are named as mega-halos. A large diffuse emission with a LLS of 5~Mpc, likely classified as a mega-halo, has also been reported in Abell~2255 \citep{Botteon2022b}. Mini-halos, on the other hand, are usually smaller (i.e., typically less than $\sim500$~kpc) and significantly brighter by an order of magnitude, as compared to radio halos \citep[e.g.][]{Cassano2008a,Murgia2009}. Mini-halos are often detected in the relaxed, cool-core clusters, whereas radio halos are predominantly observed in dynamically disturbed clusters, suggesting a link between their formation and the cluster's dynamical state.

Due to energy losses through synchrotron and inverse-Compton radiation, the radio-emitting CRes in halos and mini-halos have a lifetime of 100 Mega-years in typical ICM conditions. This suggests that the CRes are either locally \mbox{(re-)accelerated} or continuously injected throughout the ICM volume. Two main mechanisms including turbulent or hadronic processes have been proposed for the generation of these CRes. In the turbulence scenario, primary CRes are (re-)accelerated by the multi-scale turbulence induced by cluster-cluster mergers \citep[e.g.][]{Brunetti2001,Petrosian2001a,Brunetti2007a}. In the hadronic scenario, CRes are secondary products (along with gamma rays) resulting from inelastic collisions between cosmic ray protons (CRps) and thermal protons within the ICM \citep[e.g.][]{Dennison1980a,Blasi1999,Dolag2000b}. Additionally, a hybrid model suggests both primary and secondary CRes contribute to the radio emission observed in halos \citep{Brunetti2005,Brunetti2011b,Zandanel2014a,Pinzke2017a}. In the presence of magnetic fields, these primary and secondary CRes emit synchrotron emission, which is detectable in the radio band \citep[see, e.g., ][for a review of the models]{Brunetti2014}.

The pure hadronic model is disfavoured by the detection of ultra-steep spectrum radio halos (USSRHs), which exhibit spectral indices steeper than $\sim$$-1.5$, as this model typically predicts flatter spectral indices (between $-1.2$ and $-1$) \citep[i.e.,,][]{Pfrommer2004a,Brunetti2014,ZuHone2015}. In contrast, the presence of USSRHs is consistent with the turbulent model \citep{Cassano2005,Cassano2010,Brunetti2008,ZuHone2013, Brunetti2014}. Additionally, the predicted gamma rays from hadronic interactions in radio halos have not been detected by the \textit{Fermi Gamma-ray Space Telescope} \citep[][]{Jeltema2011,Brunetti2012,Zandanel2014a,Ackermann2016b,Brunetti2017,Adam2021}.  However, the hadronic mechanism has not been ruled out for mini-halos even when excluding the contribution from primary processes \citep[][]{Ignesti2020}. 

\section{The galaxy cluster RX~J1347.5-1145 
\label{sec:RXJ1347}}

RX~J1347.5-1145 (hereafter RXJ1347, $z=0.451$) is noted as the most X-ray luminous clusters, with a luminosity of $L_{X}=(6.2\pm0.6)\times10^{45}\,\textrm{erg\,\ s\ensuremath{^{-1}}}$ in the $0.1-2.4$~keV energy band \citep{Schindler1995}.
Multi-wavelength observations of the cluster have unveiled intriguing properties of this cool-core galaxy cluster during its merging process with a sub-cluster of galaxies. 

RXJ1347 has an exceptionally high mass of $(2.0\pm0.4)\times10^{15}\mathcal{M}_{\odot}$ within the $r_{200}$ radius and an overall ICM temperature of $10.0\pm0.3\,\textrm{keV}$ \citep{MyriamGitti04}. X-ray and Sunyaev–Zeldovich (SZ) observations have identified an unusual hot gas region in the SE part of the cluster, approximately $\sim150\,\textrm{kpc}$ from the cluster centre \citep{Allen2002,MyriamGitti04,Komatsu2001,Kitayama2004,Mason2010,Kitayama2016,Ueda2018}. This significant increase in the X-ray and SZ SB is thought to be due to shock-heated gas with temperature $k_{B}T_{e}>20\,{\rm keV}$, resulting from an  off-axis ongoing merger between a low-mass sub-cluster and a cool-core main cluster \cite[e.g.][]{MyriamGitti04,Mason2010,Ueda2018}. Using combined SZ (150 \& 350 GHz) and X-ray ($0.5-7.0\,{\rm keV}$ energy band) data, \citealt{Kitayama2004} estimated the temperature of the SE region to be $k_{B}T_{e}=28.5\pm7.3\,{\rm keV}$, the total electron number density to be $n_{e}=(1.49\pm0.59)\times10^{-2}\,{\rm cm^{-3}}$, and the line-of-sight length to be $240\pm183\,{\rm kpc}$. The study also found a Mach number of $2.1$ for the merger shock in the SE region. Subsequent analyses by \cite{Ota2008} using broad-band spectroscopy data from \textit{Suzaku} and by \cite{Ueda2018} using Chandra X-ray and ALMA SZ observations confirmed the presence of the hot shock gas in the SE region. The gas in the in-falling sub-cluster is stripped behind its mass peak, indicating that the SE emission is likely linked to shock-heats by the cluster-cluster merger \citep{Ueda2018}. In additional, \cite{Ueda2018} reported the detection of dipolar pattern in the X-ray emission in the core region after removing the global emission of the cluster. The authors associated this feature with the gas sloshing motion of the cool core at subsonic speed. 

The non-thermal properties of the galaxy cluster RXJ1347 remains relatively understudied despite their crucial role in understanding the magnetic field structures and the particle acceleration processes in the low-density environments of the ICM. Low-frequency radio observations with the Very Large Array (VLA) at 1.4~GHz by \citealt{Gitti2007} and the Giant Metre Radio Telescope (GMRT) at 237 and 614 MHz by \citealt{Ferrari2011} show that the cluster hosts a central diffused radio source, namely a mini-halo. The main component of the diffuse source has a diameter of 435~kpc (at 1.4~GHz). An additional diffuse emission region towards the S direction is reported in the 1.4~GHz observations, but is not detected in the 237~MHz and 614~MHz images (see Fig.~4 \& 5 in \citealt{Gitti2007} and Fig.~1 in \citealt{Ferrari2011}), extending the LLS of the cluster diffuse emission to 640~kpc  \citep{Gitti2007,Ferrari2011}. However, this southern diffuse emission might not be part of the mini-halo, but relates to the merger activity of the cluster as the sub-cluster merges to the main cluster, hosting the mini-halo, from the SW to NE direction \citep{Mason2010}. The spectral index of the mini-halo is estimated to be $\alpha_{\rm 237\,{MHz}}^{\rm 614\,{MHz}}=-0.98\pm0.05$ \citep{Ferrari2011}. Towards the SE region of the cluster, an additional component of diffuse emission is detected at a distance of $\sim125$kpc from the cluster centre. The additional diffuse radio source is spatially coincident with the region where excess X-ray and SZ emission is reported. This radio emission is proposed to originate from the merger shock via either the re-acceleration of local electrons \citep{Ferrari2011, Ensslin1998} or adiabatic gas compression \citep{Ferrari2011,Ensslin2001}. Towards the Southern region of the mini-halo, the VLA 1.4 GHz observations show the detection of an extended emission which is not visible in the GMRT 237~MHz and 610~MHz. In this paper, we aim to study the diffuse radio emission from the galaxy cluster RXJ1347 to understand the mechanisms governing the acceleration of the radio-emitting particles in the cluster.

We adopt a flat $\Lambda$CDM (Lambda Cold Dark Matter) cosmology with parameters $H_0=70\,{\rm km\,s^{-1} \,Mpc^{-1}}$, $\Omega_{\rm m}=0.3$ and $\Omega_{\Lambda}=0.7$. At the cluster redshift of $=0.451$, an angular separation of $1\arcsec$ corresponds to a physical distance of 5.767~kpc.

\section{Observations and data reduction}
\label{sec:obs_red}

\begin{table*}
	\centering
	\caption{Details of the uGMRT and MeerKAT observations of RXJ1347}
	\begin{tabular}{lccc}
		\hline\hline  
		& uGMRT 700 MHz  & uGMRT 1.26 GHz  & MeerKAT 1.28 GHz \\ \hline
		Pointing (RA, Dec)     & $13^\text{h}47^\text{m}30.6^\text{s}$, & $13^\text{h}47^\text{m}30.6^\text{s}$, & $13^\text{h}47^\text{m}30.6^\text{s}$, \\
	    & $-11^\text{h}45^\text{m}09.0^\text{s}$ & $-11^\text{h}45^\text{m}09.0^\text{s}$ & $-11^\text{h}45^\text{m}10.0^\text{s}$  \\
		Observation date       & July 2, 2021 & June 8, 2021; July 6, 2021 & September 24, 2022 \\
		On-source time (hr)    & 4.5 & 9.5  & 8.3 \\
		Freq. coverage (MHz)   & $550 - 950$ & $1060 - 1460$ & $890 - 1668$  \\
		Usable bandwidth (MHz) & 300 & 394 & 778 \\
        Central frequency (MHz) & 700 & 1260 & 1279 \\
		Channel width (kHz)    & 195 & 195 & 209   \\
		Integration time (s)   & 8 & 16, 8 & 8 \\
		Correlation            & RR, LL, RL, LR & RR, LL, RL, LR & XX, YY, XY, YX    \\
		Number of antennas     & 28  & 30 & 62 \\ \hline\hline
	\end{tabular}  
	\label{tab:obs_radio}
\end{table*}

\begin{table*}
\centering
	\caption{Parameters of the radio imaging of RXJ1347.}
	\begin{tabular}{lcccccc}
		\hline\hline
		Data & $uv$-range & $\mathtt{Robust}^a$ & $\theta_\text{\tiny FWHM}$ & $\sigma_{\text{\tiny rms}}$ & Fig.  \\
		& (k$\lambda$) & ($\mathtt{outertaper}$) & ($\arcsec\times\arcsec$, $PA^b$) & ($\mu\text{Jy}\,\text{beam}^{-1}$) & &  \\ \hline
		uGMRT 700 MHz & $0.12-67$ &$0$ & $5.8\times3.6$ ($24^\circ$) & 20 & \ref{fig:uGMRT_MeerKAT} \\
        & $0.12-67$ &$0$ ($5\arcsec$) & $7.7\times5.5$ ($41^\circ$) & 32 & \ref{fig:uGMRT_MeerKAT}$^d$ \\
		& $0.12-41$ & $0$ ($5\arcsec$) & $9\times9^c$ & 51 & \ref{fig:spx}$^d$ \\ \hline
       uGMRT 1.26 GHz & $0.21-120$ & $0$ & $3.0\times2.2$ ($80^\circ$) & 18 & \ref{fig:uGMRT_MeerKAT} \\
            & $0.21-120$ & $0$ ($10\arcsec$) & $10.2\times8.5$ ($72^\circ$) & 121 & \ref{fig:uGMRT_MeerKAT}$^d$ \\
                \hline
        MeerKAT 1.28 GHz & $0.06-41$ & $-0.5$ ($5\arcsec$) & $7.5\times7.0$ ($-16^\circ$) & 10 & \ref{fig:uGMRT_MeerKAT}  \\
		& $0.06-41$ & $0.25$ & $9.5\times8.9$ ($-23^\circ$)& 10 & \ref{fig:HST_MeerKAT} \\    
        & $0.06-41$ & $0.25$ ($10\arcsec$) & $13.7\times13.2$ ($-56^\circ$)& 15 & \ref{fig:uGMRT_MeerKAT}$^d$ \\ 
        & $0.12-41$ & $0$ & $9\times9^c$ & 10 & \ref{fig:spx}$^d$ \\  \hline \hline
	\end{tabular}\\	
	Notes: $^a$: Briggs weighting of visibilities; $^b$: position angle (PA) of the elliptical synthesized beam; $^c$: smoothed to the targeted resolution; $^d$: BCG radio emission subtracted from the data.
	\label{tab:imaging_para}
\end{table*}

\subsection{Radio data}
\label{sec:radio}
\subsubsection{MeerKAT 1.28~GHz}
\label{sec:meerkat_data}

MeerKAT L-Band observations of RXJ1347 were conducted on September 24, 2022 with a total on-source observing time of 8.3 hours. The cluster field was observed using full polarization settings, covering a frequency range from 890~MHz to 1668~MHz, with a channel width of 209~kHz. Compact radio sources including J0408-6545, J1337-1257, and J1939-6342 were observed for calibration purposes. More details on the MeerKAT 1.28~GHz observations are provided in Table~\ref{tab:obs_radio}.

The MeerKAT data was calibrated using the SARAO Science Data Processor (SDP) pipelines\footnote{\url{https://skaafrica.atlassian.net/wiki/spaces/ESDKB/pages/338723406/SDP+pipelines+overview}}. The initial calibration pipeline was first used to solve and apply  gain, bandpass and phase corrections, and to remove radio frequency interference (RFI). For the flux scale and bandpass calibration, the radio galaxies J0408-6545 and J1939-6342 with flux densities of 15.709~Jy and 15.007~Jy, respectively, were utilized. Subsequently, the data was processed with the MeerKAT continuum pipeline, which performs self-calibration to correct for wide-field effects across multiple sub-regions (facets) of the target field. The calibrated data was then deconvolved to create continuum intensity images using the Common Astronomy Software Package\footnote{\url{https://casa.nrao.edu/}} (CASA). It is worth noting that the attenuation of the MeerKAT primary beam in the outskirts of RXJ1347 is minimal due to the cluster's small angular size (i.e., $3\arcmin$). The telescope sensitivity in these outskirts remains up to 99~percent of that in the pointing centre. Nonetheless, for completeness, we applied the primary beam correction to the final images using the MeerKAT primary beam model described by \cite{Asad2021}. To reveal multiple emission scales, we create multiple images with sets of parameters given in Tab.~\ref{tab:imaging_para}.

\subsubsection{uGMRT 700 MHz and 1.26~GHz}
\label{sec:gmrt_data}

RXJ1347 was observed with the uGMRT at 700~MHz (Band 4) and 1.26~GHz (Band 5) for a total duration of 4.5 and 9.5 hours, respectively. The observations have an usable bandwidth of 300 MHz and 394 MHz with the central frequencies at 700 MHz and 1260 MHz, respectively. More details are presented in Table.~\ref{tab:obs_radio}.

The uGMRT data were calibrated using the Source Peeling and Atmospheric Modelling \citep[\texttt{SPAM};][]{Intema2009a} pipeline, which provides high-fidelity images for extended radio emission for interferometric data from the uGMRT. This semi-automatic pipeline corrects for both direction-independent and direction-dependent effects that can distort the propagation of radio signals from astronomical sources to the telescope receivers, given the telescope's wide field of view (FWHM of $40'$ at 700 MHz and $22'$ at 1.26~GHz). 
To calibrate the wide-band uGMRT data, the data were split into six sub-bands, each with a bandwidth of 33.3~MHz, to minimise the calibration errors caused by the non-uniform flux density of sources across the broader bandwidth (300 MHz for Band 4 and 394~MHz for Band 5). Each sub-band data set was processed separately with \texttt{SPAM}, using 3C~295 as the primary flux calibrator that is set to the flux scale of \cite{Scaife2012}. Initial phase calibration was performed on the wide-field dataset, followed by multiple iterations of amplitude and phase direction-dependent calibration on various sub-fields (facets) of the target field. The direction-dependent calibration solutions were used to derive the ionospheric corrections that were applied to the data. We note that we skip the ionospheric corrections for some of the uGMRT 1.26 GHz sub-datasets due to the minor effects of the ionosphere on the radio signal at this frequency range (i.e., using the option allow\_selfcal\_skip = True). The calibrated sub-band datasets were then performed a phase and amplitude self-calibration to minimise the difference due to the calibration of the sub-datasets. The self-calibrated data were concatenated and deconvolved to produce total intensity images of the target field. Finally, primary beam correction was applied using an uGMRT primary beam correction script\footnote{\url{https://github.com/ruta-k/uGMRTprimarybeam-CASA6}} to obtain the final images. The imaging parameters are shown in Tab.~\ref{tab:imaging_para}. 

\subsubsection{Discrete source subtraction}
\label{sec:source_subtraction}

\begin{table}
\centering
	\caption{The flux density of the central radio galaxy}
	\begin{tabular}{lcc}
		\hline\hline
		Data & Flux density (mJy) & Ref. \\ \hline
		GMRT 237 MHz & $55\pm4$ & \cite{Ferrari2011} \\  
        GMRT 614 MHz & $32\pm2$ & \cite{Ferrari2011}  \\
        uGMRT 700 MHz & $35.6\pm1.8$ & This work \\
        uGMRT 1.26 GHz & $35.5\pm1.8$ & This work \\
        MeerKAT 1.28 GHz & $31.7\pm1.6$ & This work \\
        VLA 1.4 GHz & $29.8\pm0.3$ & \cite{Gitti2007} \\
        \hline \hline
	\end{tabular}\\	
	\label{tab:central_radio}
\end{table}

\begin{figure}
\centering 
    \includegraphics[width=1\columnwidth]{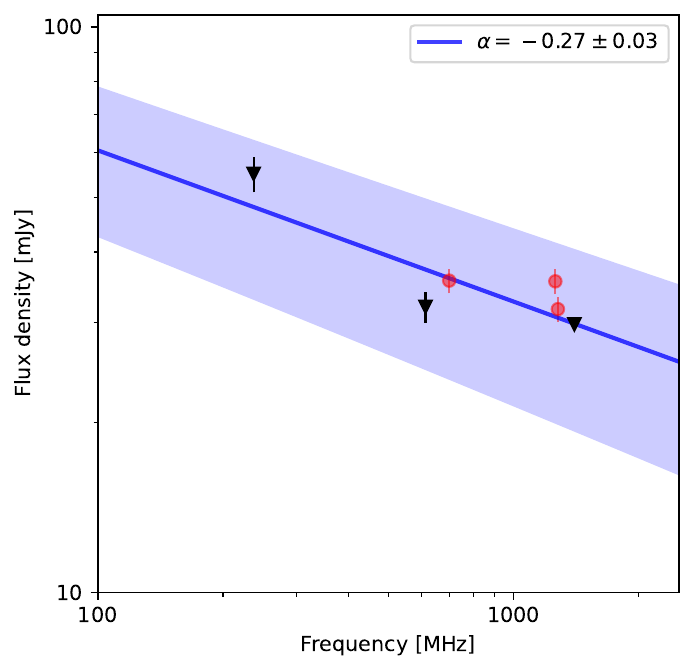}
    \caption{Integrated spectrum for the central radio galaxy. The data points measured in this work are marked in red. The light blue region shows the $1\sigma$ uncertainty of the best-fit line. }
    \label{fig:BCG_int}
\end{figure}

Radio emission from galaxies was observed in the central region of RXJ1347 \citep[see Fig.~2 and 3 in][]{Gitti2007}, with a prominent source originating from the central BCG \citep[i.e., the BCG A in Fig.~7 of ][]{Mason2010}. To remove these discrete radio sources, primarily focusing on the removal of the central radio galaxy, we create high-resolution models using WSClean \citep{Offringa2017}. To filter out the cluster's diffuse emission, we generated images with $uv$ minimums between 5~$k\lambda$ and 19~$k\lambda$ (in increment of 2~$k\lambda$) and applied \textit{Briggs} weighting with a robust parameter of $-2$. The models were then subtracted from the $uv$ data.

In Fig.~\ref{fig:BCG_sub_uv_mins}, we show the MeerKAT images of the cluster central region before and after the subtraction, which show the detection of the central radio galaxy as a point-like source. This is consistent with the VLA 1.4~GHz observations \citep{Gitti2007} and the Gaussian model of the source at 614~MHz \citep{Ferrari2011}. The images also show the present of the diffuse emission from the ICM which becomes less prominent in the images with higher $uv$ cuts ($uv_{\rm min}\geq13~k\lambda$). The flux densities of the central radio galaxy measured in these images remain relatively constant (see  Fig.~\ref{fig:BCG_sub_uv_mins}). Its flux density, obtained from the 13~$k\lambda$ $uv$ cut-off image where the radio galaxy begins to separate from the diffuse emission, is $31.7\pm1.6$~mJy. 

Similarly, we measured the flux density of the central radio galaxy from the uGMRT 700~MHz and 1.26~GHz images with a $uv$ minimum cut of 13~$k\lambda$. Our measurements, along with those reported by \cite{Gitti2007} and \cite{Ferrari2011}, are given in Tab.~\ref{tab:central_radio} and are fitted with an exponential function. The spectrum of the radio galaxy is plotted in Fig.~\ref{fig:BCG_int}. Our measurements align with the previous ones and fall within the 1$\sigma$ uncertainty range of the best-fit power-law line, indicating that the central radio galaxy is effectively modelled and subtracted from the data.

\subsection{\textit{Chandra} X-ray data}
\label{sec:chandra_data}

RXJ1347 was observed multiple times with the Advanced CCD Imaging Spectrometer (ACIS), totalling 234~ks of observing time. Details of the observations are given in Table~\ref{tab:obs_xray}. We reprocessed the archival data using the \texttt{ClusterPyXT} package \citep{Alden2019} that makes use of the \textit{Chandra} Interactive Analysis of Observations (\texttt{CIAO}) package and the \textit{Chandra} Calibration Database (CalDB; version 4.9.2). During the reprocessing, compact sources in the observing field were visually identified and are masked out. No compact source is found within the radius of 500~kpc from the cluster centre (i.e., the X-ray SB peak). The observations were then merged to generate a combined image of the cluster field. For more details on the processing, we refer to \citep{Alden2019}.

\begin{table}
	\centering
	\caption{Details of the \textit{Chandra} observations of RXJ1347}
	\begin{tabular}{lccc}
		\hline\hline
		ObsID & Instrument & Mode & Exposure time \\
            &              &       &  (ks)        \\ \hline     
        506  & ACIS-S & VFAINT & 8.93     \\
        507  & ACIS-S & FAINT  &  10.00 	   \\
        3592  & ACIS-I & VFAINT &   57.71   \\
        13516  & ACIS-I & VFAINT &  39.56   \\
        13999  & ACIS-I & VFAINT &  54.35   \\
        14407  & ACIS-I & VFAINT &  63.24    \\ \hline
         & & Total &  234    \\ 
        \hline\hline
	\end{tabular}
	\label{tab:obs_xray}
\end{table}

\section{Results}
\label{sec:res}

In Figs.~\ref{fig:HST_MeerKAT} and \ref{fig:uGMRT_MeerKAT}, we present the multi-wavelength radio maps of the galaxy cluster RXJ1347, taken with the MeerKAT at 1.28~GHz and the uGMRT at 1.26~GHz and 700~MHz. Fig.~\ref{fig:Chandra_MeerKAT} shows the \textit{Chandra} X-ray map overlaid with the MeerKAT radio emission contours. The following sub-sections detail the observed properties of the cluster's diffuse radio emission and its correlation with the thermal X-ray diffuse emission.

\subsection{Diffuse radio emission}
\label{sec:res_diff_emission}

\begin{figure*}
\centering
    \includegraphics[width=0.7\textwidth]{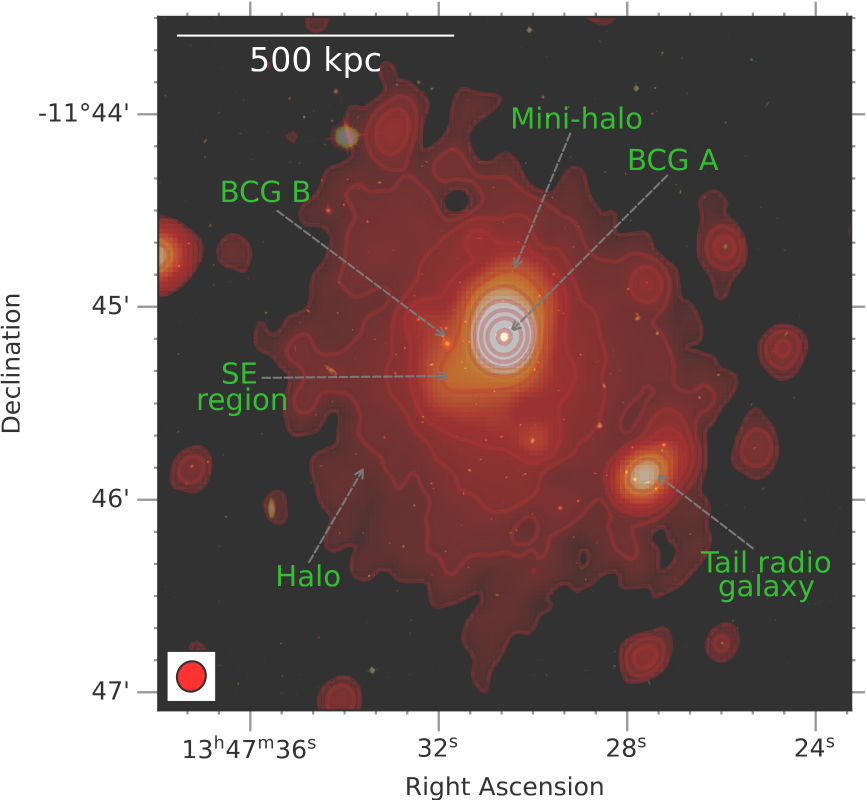}
    \caption{MeerKAT 1.28~GHz image of RXJ1347 overlaid with the HST RGB-band colour image. The radio emission from the central BCG has not been subtracted in this image. The radio contour starts from $3\sigma$, where $\sigma=10\,{\rm \mu Jy\, beam^{-1}}$, and are subsequently spaced by a factor of 2. The synthesis beam of  $9.5\arcsec\times8.9\arcsec$ (position angle of $-23$) for the radio data is shown in the bottom left corner.}
    \label{fig:HST_MeerKAT}
\end{figure*}

\begin{figure*}
\centering
    \includegraphics[width=0.33\textwidth]{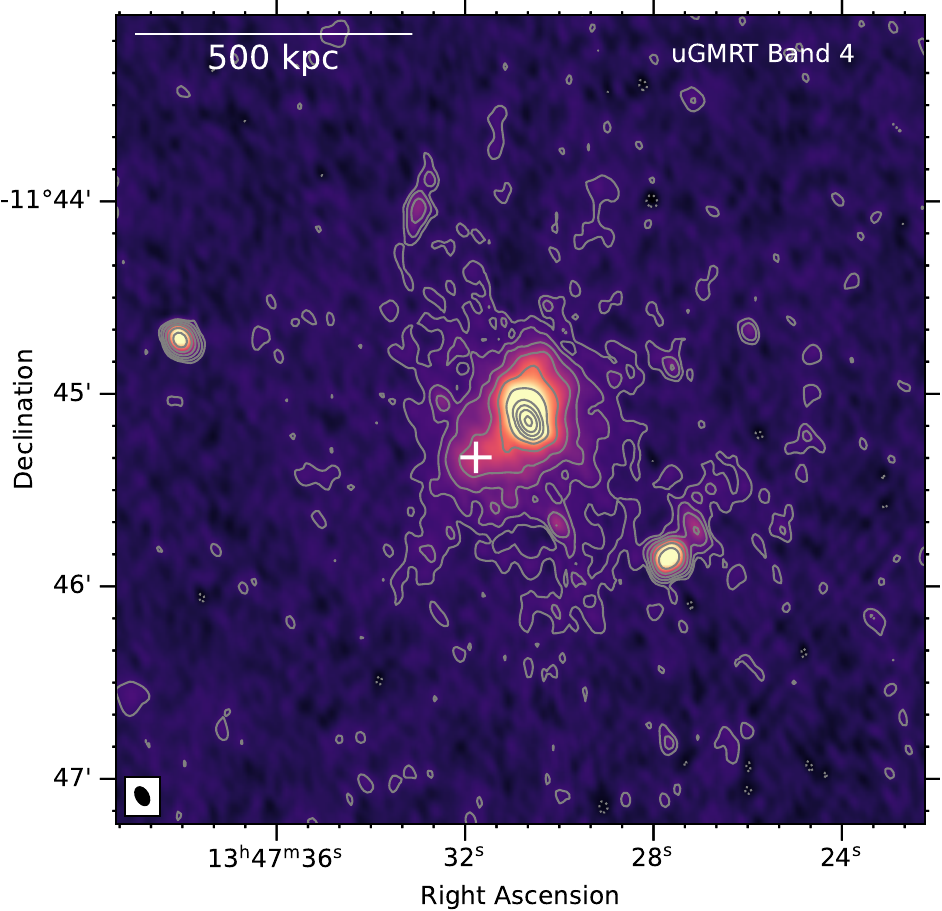} \hfil
    \includegraphics[width=0.33\textwidth]{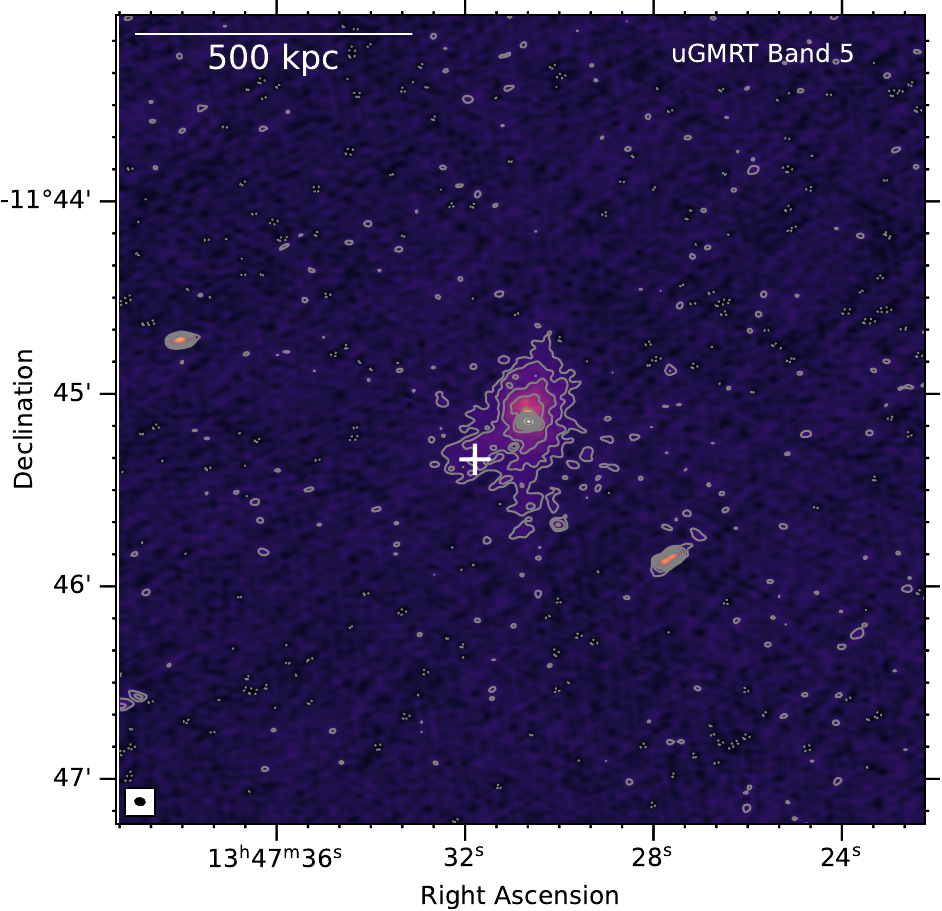} \hfil
    \includegraphics[width=0.33\textwidth]{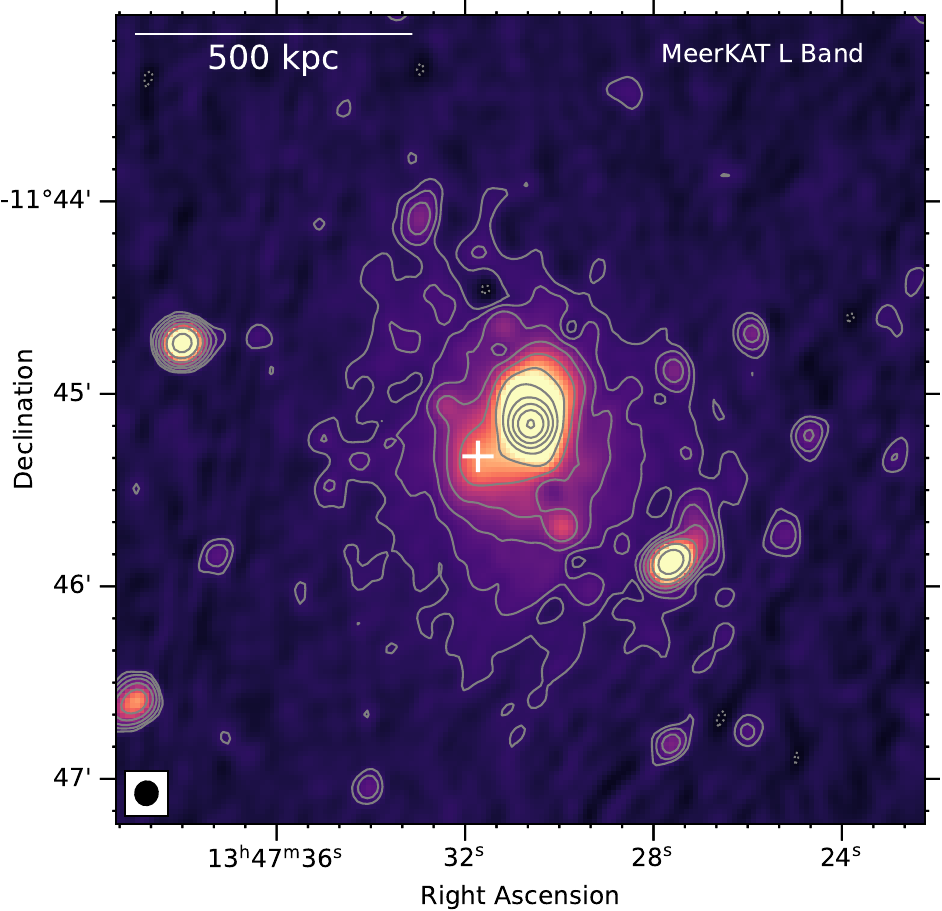} \\
    \includegraphics[width=0.33\textwidth]{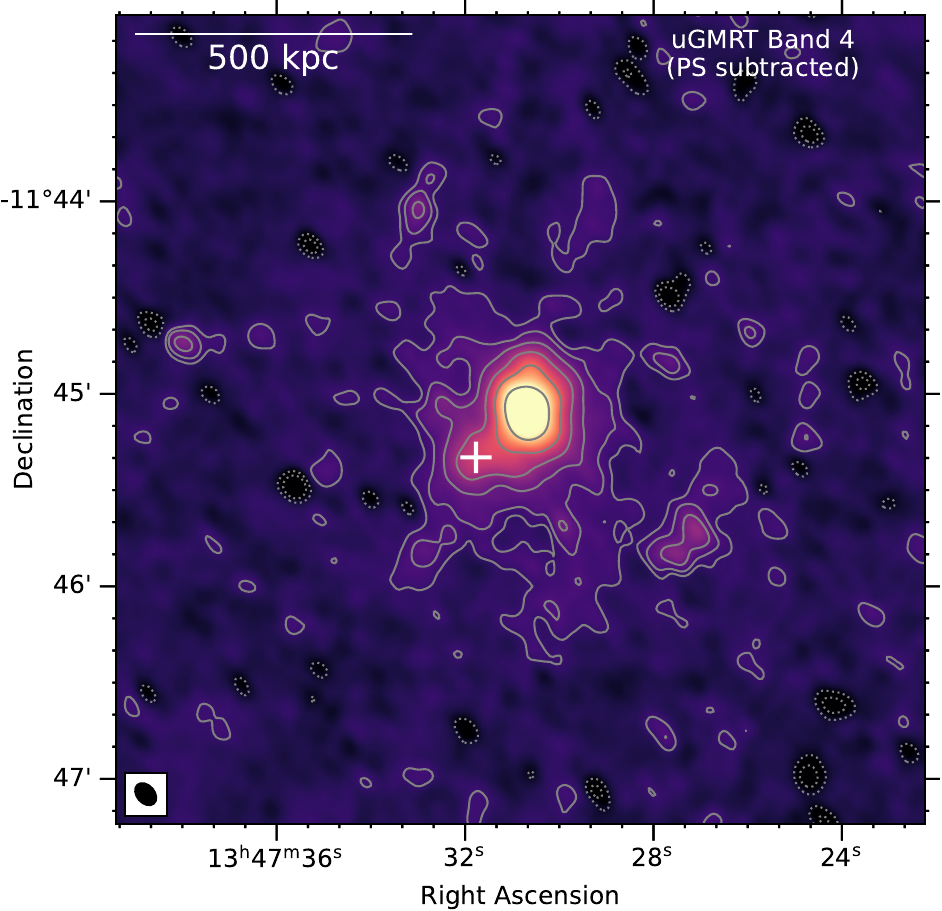} \hfil
    \includegraphics[width=0.33\textwidth]{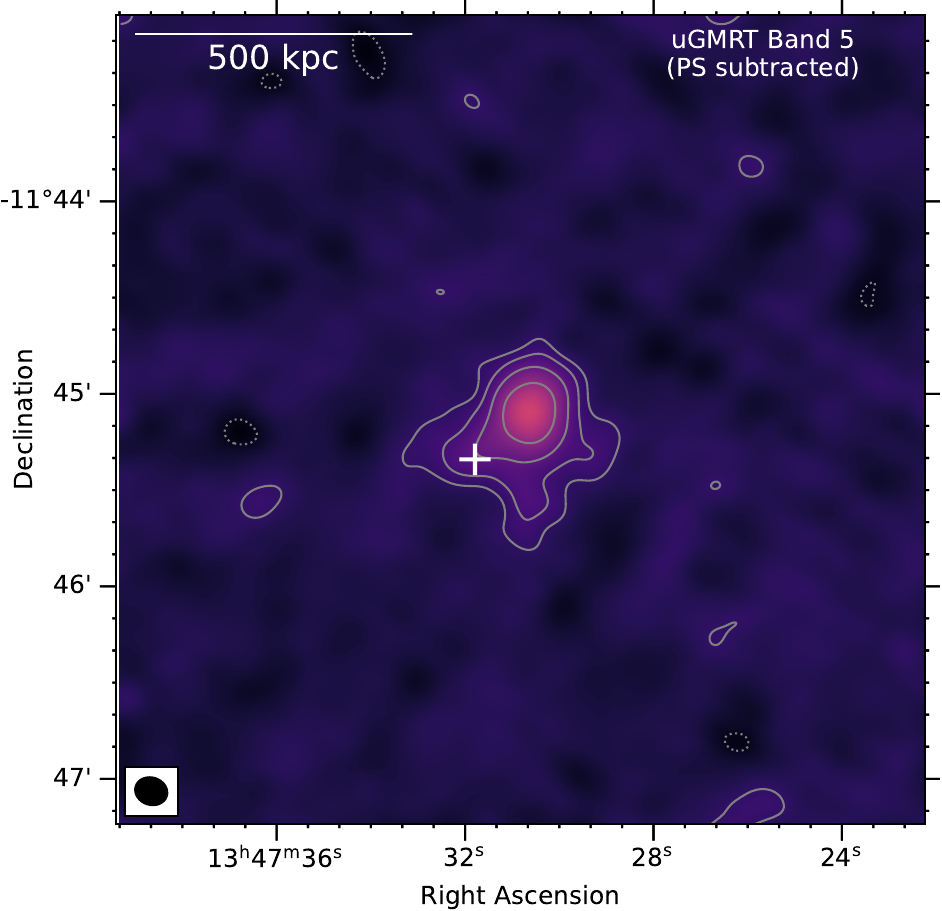} \hfil
    \includegraphics[width=0.33\textwidth]{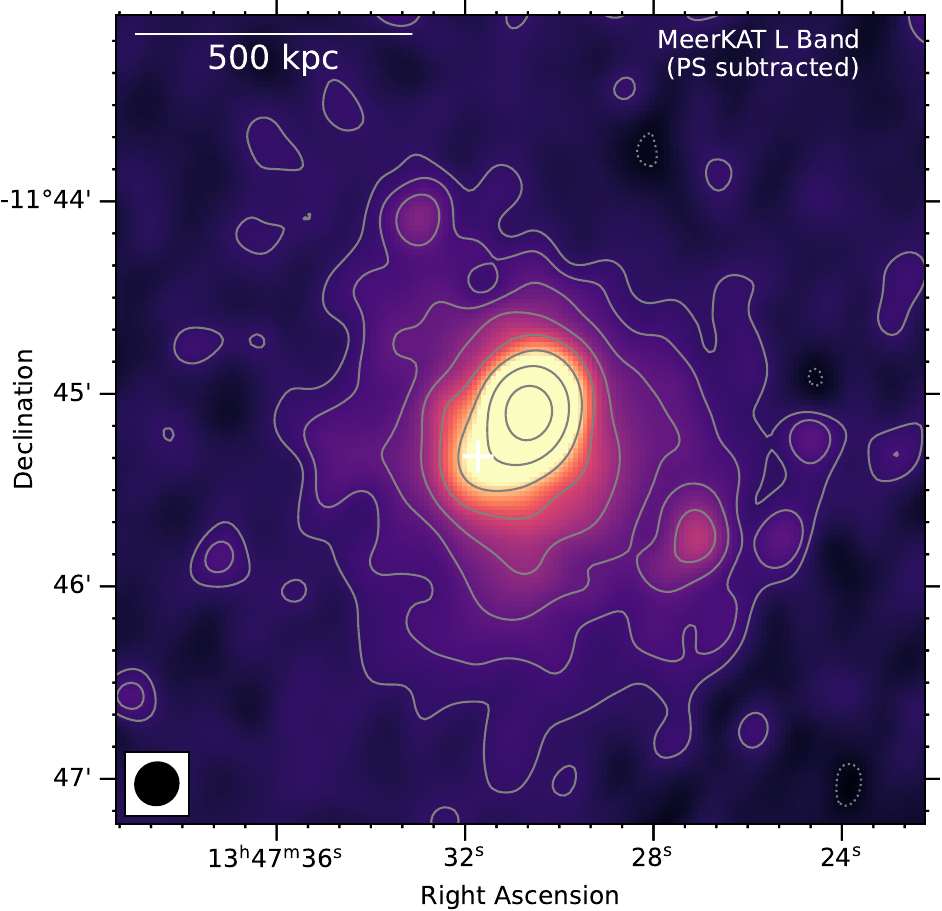} \\
    \caption{Radio intensity maps of RXJ1347 at 700~MHz (left column), 1.26~GHz (middle column), and 1.28~GHz (right column). The discrete sources including the central BCG are removed in the second row. The first thick contours are drawn at $3 \sigma$, and the subsequent contours are multiplied by a factor of 2. The thin contour levels at $2 \sigma$. Here $\sigma$ is $20\,{\rm \mu Jy\,beam^{-1}}$ (and $32\,{\rm \mu Jy\,beam^{-1}}$) at 700~MHz, $18\,{\rm \mu Jy\,beam^{-1}}$ (and $121\,{\rm \mu Jy\,beam^{-1}}$) at 1.26~GHz, and $10\,{\rm \mu Jy\,beam^{-1}}$ (and $15\,{\rm \mu Jy\,beam^{-1}}$) at 1.28~GHz for the first (and second) row images. The SE region where excess radio and X-ray emission is observed is marked with the plus (+) sign. The beam is displayed in the lower left-hand corner in each panel. }
    \label{fig:uGMRT_MeerKAT}
\end{figure*}

\begin{figure}
\centering
\includegraphics[width=1\columnwidth]{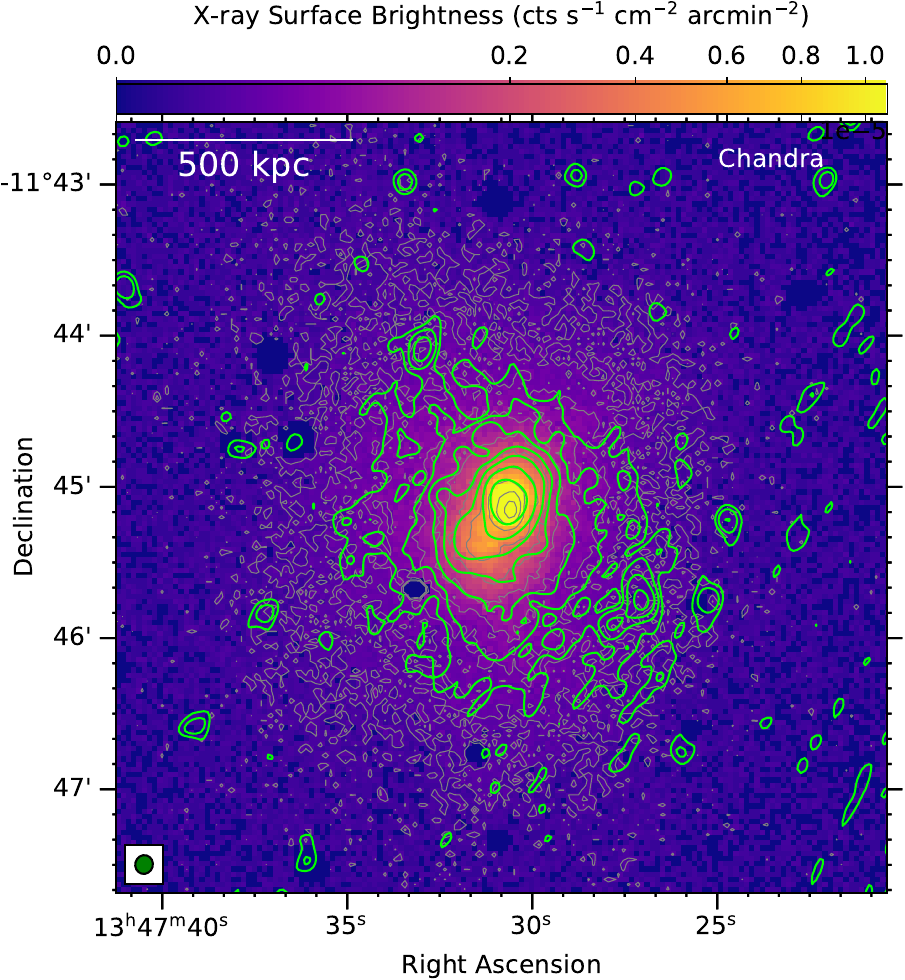}
  \caption{\textit{Chandra} X-ray image overlaid with its gray contours. The MeerKAT 1.28 GHz (green) contours are also displayed. The discrete sources are removed from the \textit{Chandra} and MeerKAT data. 
          }
     \label{fig:Chandra_MeerKAT}
\end{figure}

The MeerKAT 1.28 GHz images in Figs.~\ref{fig:HST_MeerKAT} and \ref{fig:uGMRT_MeerKAT} (right panels) shows the extended radio emission in the central region of the cluster, known as a mini-halo \citep{Gitti2007,Ferrari2011}. With our sensitive observations, the diffuse emission is detected further to the cluster periphery that is beyond the reported size (i.e., LLS$=640$~kpc). The largest linear projected size of the diffuse radio emission in the MeerKAT image in Fig.\ref{fig:uGMRT_MeerKAT} (bottom right panel) is 1~Mpc. Located in the centre of the cool-core main cluster is the BCG that hosts a radio-loud AGN galaxy \citep{Gitti2007,Mason2010}. As seen in Fig.~\ref{fig:HST_MeerKAT}, approximately 100~kpc to the eastern side from the main BCG is a radio-quiet AGN galaxy (i.e., the second BCG) belonging to the sub-cluster merging from the SW direction towards the NE \citep{Gitti2007,Mason2010}. An excess of diffuse radio emission is detected in the region about 125~kpc towards the SE from the centre of the main cluster, where excess X-ray emission was also reported \citep[e.g.][]{Ota2008}. 

Fig.~\ref{fig:uGMRT_MeerKAT} displays the uGMRT images (left and middle panels) at the frequency of 700~MHz and 1.26~GHz. In the second row, the discrete sources including the central AGN radio galaxy are removed.  In the uGMRT images, diffuse radio emission is detected in the central region of the cluster and has a largest linear size of 550~kpc and 380~kpc, respectively. The source projected size in the uGMRT 1.26~GHz image is half of that in the equivalent frequency, resolution MeerKAT 1.28~GHz image, which is likely due to the missing of the short baselines and/or the low sensitivity of the uGMRT observations. Towards the SE region of the cluster, additional diffuse emission is also seen in the uGMRT images (i.e., marked with the plus sign in Fig.~\ref{fig:uGMRT_MeerKAT}). This is coincident with that in the MeerKAT image and with the reports in literature \citep{Gitti2007,Ferrari2011}.

\begin{figure*}
\centering
    \includegraphics[width=0.49\textwidth]{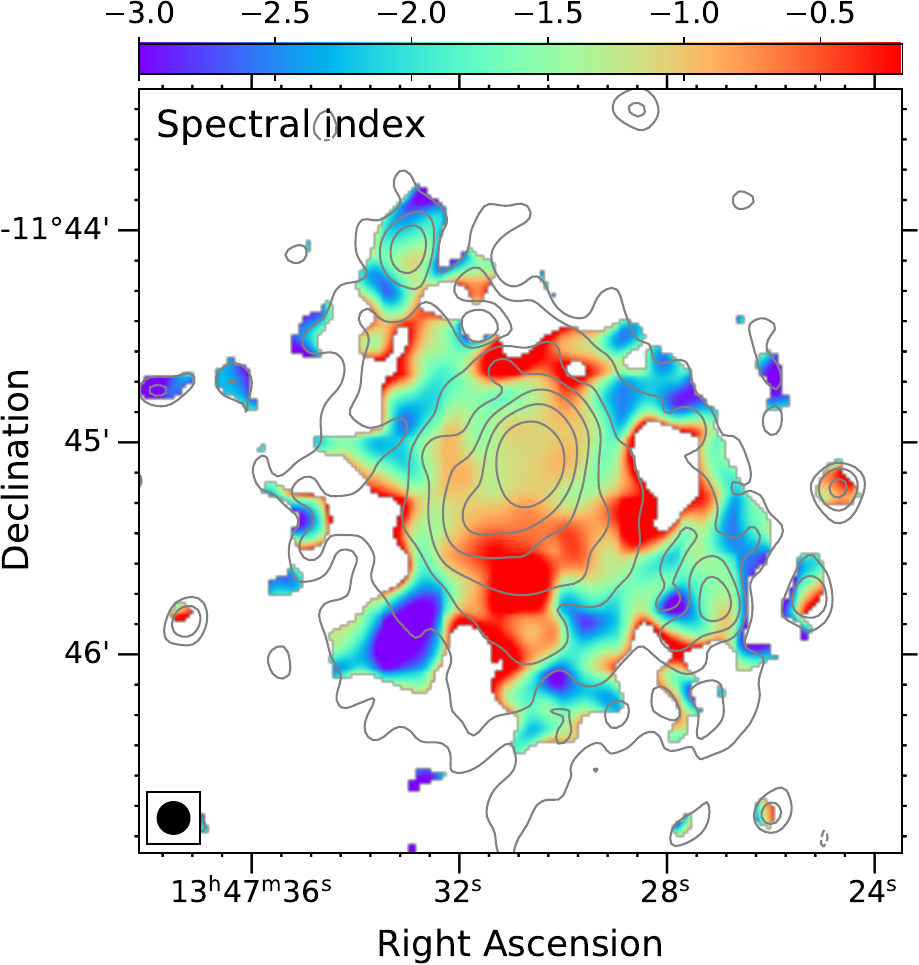}  \hfil
    \includegraphics[width=0.49\textwidth]{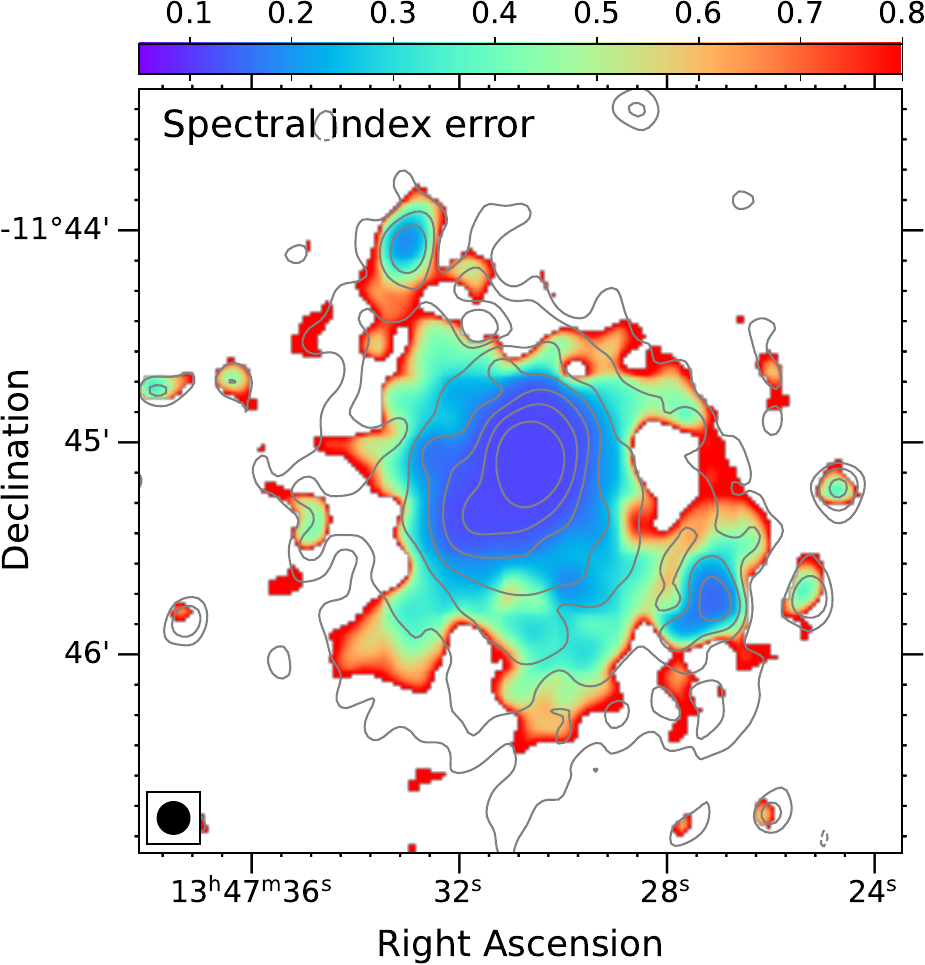} 
    \caption{Left: The distribution of spectral indices between 700~MHz and 1.28~GHz. The overlaid contours are drawn from the corresponding MeerKAT 1.28~GHz map. The contours starts from $3\sigma$, and are spaced by a factor of 2. Right: The associated spectral index error.}
    \label{fig:spx}
\end{figure*}

In Fig.~\ref{fig:spx} we show the spectral index map that is made with the uGMRT 700~MHz and MeerKAT 1.28~GHz datasets. To ensure equivalent sensitivity to diffuse emission across different scales, the uGMRT and MeerKAT intensity images are created with a common $uv$ range between $0.120\,k\lambda$ and $41\,k\lambda$ (see Table~\ref{tab:imaging_para}). The spectral index map reveals patchy distribution in the cluster central region, but it remains relatively flat with a mean value of $\alpha=-1.10$ and a standard deviation of $0.08$. This variation is still within the typical spectral index error of $\sim$$0.12$ in the region. In the cluster outskirts, the spectral index becomes steeper, down to  $\sim$$-3$. Behind the SE shock-heated region, the spectral index does not indicate any clear spatial gradient trends.

\subsection{Radial SB profiles}
\label{sec:res_sb_profiles}

\begin{figure*}
\centering 
    \includegraphics[width=0.47\textwidth]{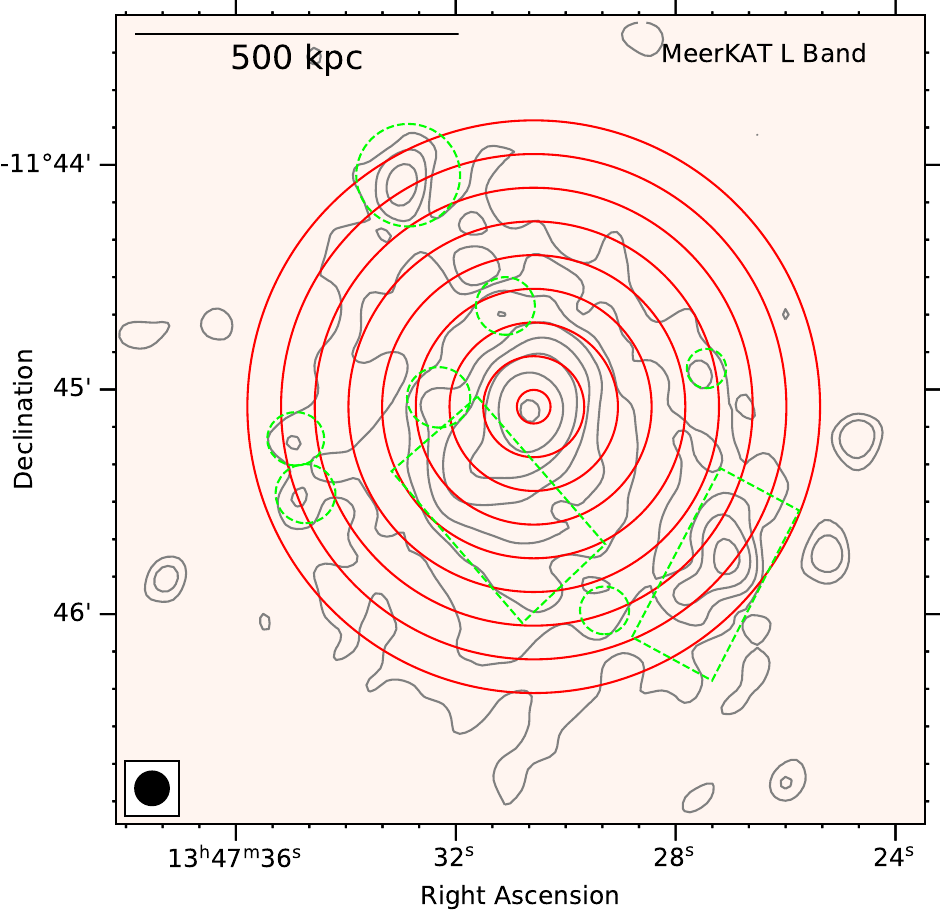}  \hfil 
    \includegraphics[width=0.49\textwidth]{RXJ1347_MeerKAT_Double_Exp_Prof.pdf} 
    \caption{Radial SB profiles. Left: Annular regions (in red) where SB is extracted. The widths of the annulus are equal to $9\arcsec$. The radio emission from the discrete sources are masked with the green regions and are not used in the analysis. Right: radio radial profile and the best-fit model.        }
    \label{fig:profile_r}
\end{figure*}

The MeerKAT~1.28 GHz discrete-source-subtracted image in Fig.~\ref{fig:uGMRT_MeerKAT} (right panels) indicates possible existence of different radial gradients in the SB of the diffuse radio emission. To further examine this, we make the radial SB profile using the MeerKAT 1.28~GHz data. The SB is extracted within the annuluses centred on the peak emission of the radio image (see Fig.~\ref{fig:profile_r}, left panel). The width of the annulus is set at $9\arcsec$ which is equal to the size of the synthesized beam of the image. To avoid the contamination from imperfect source subtraction, we masked out the radio emission from discrete sources and the SE excess diffuse emission. In  Fig.~\ref{fig:profile_r} (right panel) we show the extracted radio SB profile.

Diffuse radio emission in the cluster central regions, known as mini-halos and halos, generally follows the exponential functions of the form,
\begin{equation}
I (r) = I_0\,\exp{(-r/r_e)},
\label{eq:radial_profile}
\end{equation}
where $I_0$ is the peak intensity and $r_e$ is the $e$-folding radius that qualifies how fast the SB decreases as a function of radius \citep[e.g.][]{Murgia2009}. We fit the radio SB profile of RXJ1347 to a double exponential function,  
\begin{equation}
\label{eq:double_expo}
I (r) = I_{\rm inner} (r) + I_{\rm outer} (r),
\end{equation}
where $I_{\rm inner}$ and $I_{\rm outer}$, given individually in Eq.~\ref{eq:radial_profile}, are the mean radial intensity for the inner and outer components of the diffuse emission. We obtain the best-fit parameters for the double exponential model in Table.~\ref{tab:radial_fitting_rx}. The best-fit lines for the SB profile are presented in Fig.~\ref{fig:profile_r} (right panel). The peak intensity for the inner component is a factor of 20 times brighter than the outer one, $I_{\rm 0,\,inner} \approx 20\,I_{\rm 0,\,outer}$.  The inner $e$-folding radius is roughly seven times smaller than that in the outer component, $r_{e, {\rm inner}} \approx \nicefrac{r_{e, {\rm outer}}}{5}$. 

Following \cite{Murgia2009}, we use the best-fit parameters to estimate the flux densities for the diffuse sources by integrating the flux densities (i.e., Eq.~\ref{eq:radial_profile}) to a radius of $3\,r_e$. These flux densities are equal to 80 percent of the total flux density when integrating to infinity. We obtain flux densities of $20.4\pm1.7\,{\rm mJy}$ and $21.8\pm11.0\,{\rm mJy}$ at 1.28~GHz for the inner and outer components, respectively. These correspond to the $k$-corrected radio power of $(15.9\pm1.7)\times10^{24}\,{\rm W\,Hz ^{-1}}$ and $(19.0\pm10.0)\times10^{24}\,{\rm W\,Hz ^{-1}}$. Here we use the spectral indices of $-1.1\pm0.2$ and $-1.4\pm0.2$ for the inner and outer regions, respectively. 

\begin{table}
	\centering
	\caption{Radial SB profile fitting results for the MeerKAT 1.28 GHz data}  
	\begin{tabular}{lcc}
		\hline\hline
		   Double exp. & & \\ \hline
		$I_{0, MH}$ (mJy~beam$^{-1}$) & $6.1\pm0.8$ \\
        $r_{e, MH}$ (kpc) & $43\pm5$ \\
        $I_{0, H}$ (mJy~beam$^{-1}$) & $0.3\pm0.2$  \\
        $r_{e, H}$ (kpc) & $204\pm58$ \\
        $\chi_{\rm reduced}^2$ & 10 \\ 
        \hline\hline
	\end{tabular}
	\label{tab:radial_fitting_rx}
\end{table}

\subsection{Spatial radio -- X-ray SB correlation}
\label{sec:res_radio_xray_corr}

Radio and X-ray diffuse emission from cluster central regions are known to correlate, thus probing the connection between the non-thermal (i.e., relativistic electrons and magnetic fields) and thermal components \citep[e.g.][]{Govoni2001c,Ignesti2020,Botteon2020a,Biava2021,Riseley2022a,Bonafede2022,Lusetti2023}. This correlation is generally described with a power-law relation of the form\footnote{$\log_{10}$ is used throughout this paper.},
\begin{equation}
    \log {I_R} = a + b \log {I_X}, 
    \label{eq:radio_xray}
\end{equation}
where $a$ and $b$ are free parameters. The slope $b$ indicates how tight the radio and X-ray emission is correlated. The radio SB reduces faster (or slower) compared with the X-ray SB for larger (or smaller) value of $b$. The radio and X-ray SB is linearly related in case of $b=1$. The relation is super-linear for $b>1$; and vice versa, it is a sub-linear relation for $b<1$. 

\begin{figure*}[!ht]
    \centering
    \includegraphics[width=0.32\textwidth]{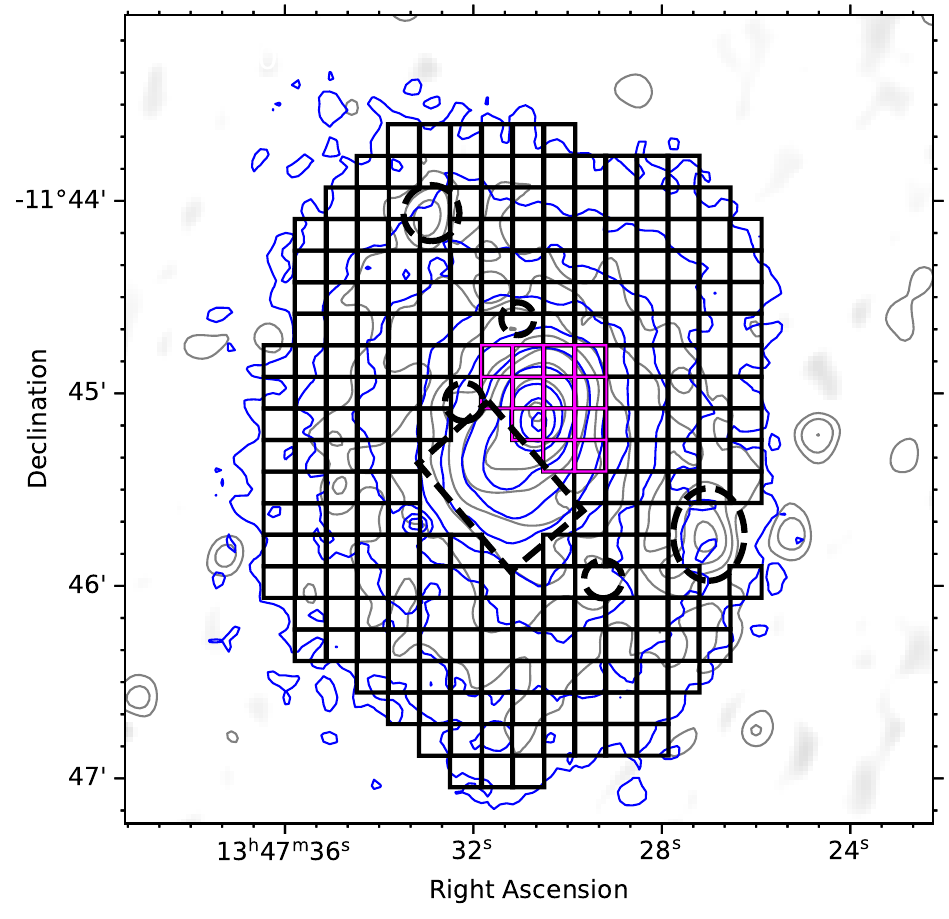}  \hfil
    \includegraphics[width=0.33\textwidth]{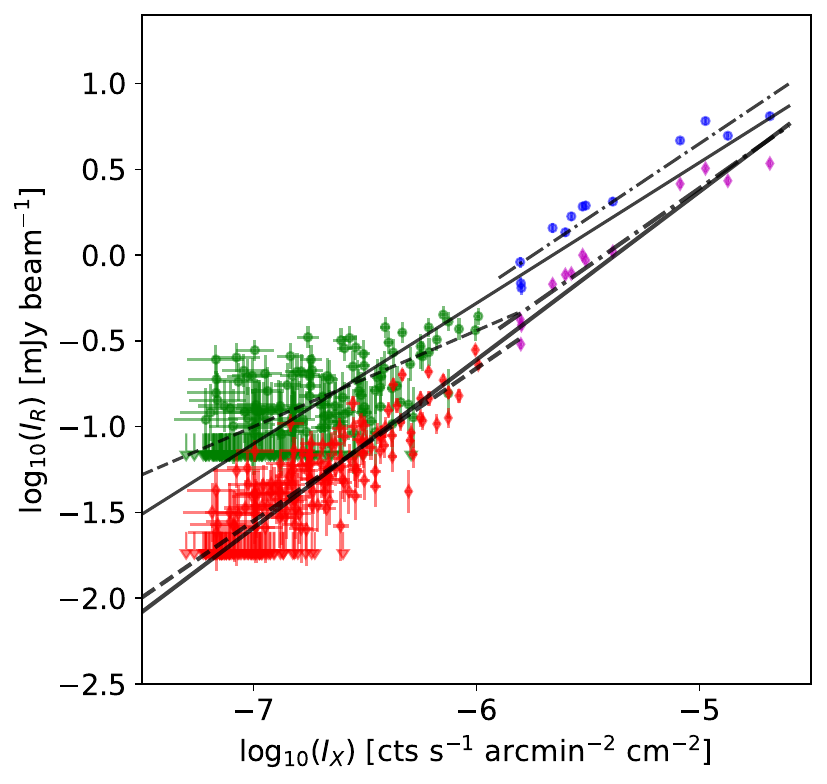} \hfil 
    \includegraphics[width=0.32\textwidth]{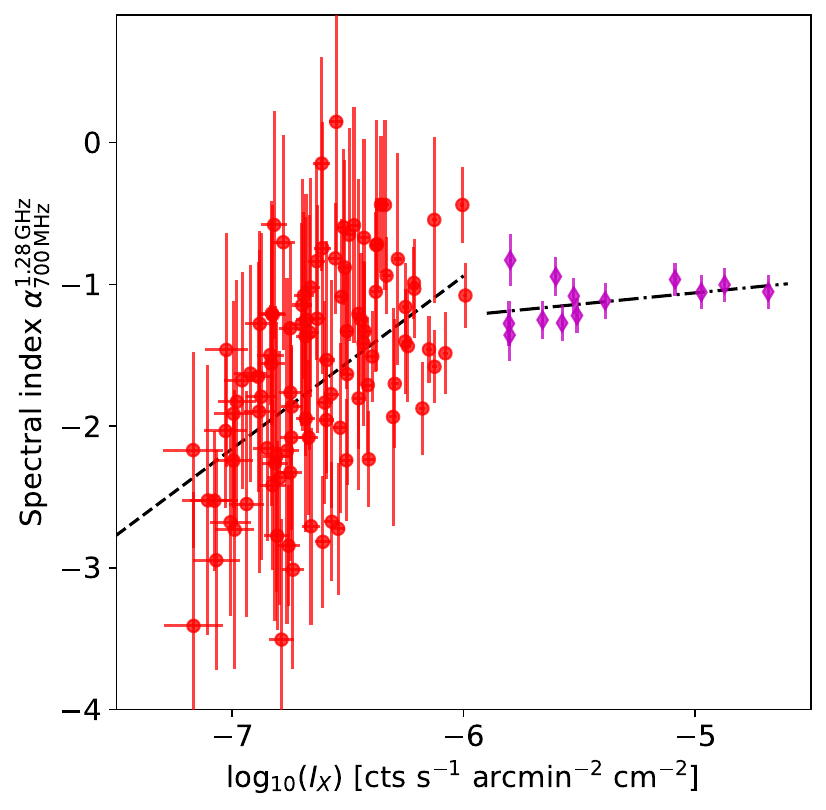}
    \caption{Spatial radio -- X-ray SB correlation. Left: The box regions, within the $3\sigma$ \textit{Chandra} (blue) contour, where the SB is extracted. The first \textit{Chandra} (blue) and MeerKAT (black) contours are levelled at $3\sigma$, where $\sigma=2\times10^{-8}$ and $\sigma=12\,{\rm \mu Jy\,beam^{-1}}$, respectively. The next contours are spaced by a factor of 2. Discrete sources are masked and indicated by black dashed lines. Middle: The radio -- X-ray correlation for the uGMRT 700 MHz data (i.e., green and blue points) and the MeerKAT 1.28 GHz data (i.e., red and magenta points) with the \textit{Chandra} data. The thin and thick solid lines are the best-fit lines for the uGMRT and MeerKAT data, respectively. The dashed-dotted and dashed lines are obtained with the data in the inner  (magenta) and outer regions, respectively. Right: The correlation between the spectral index and the X-ray SB. The data points with the best-fit (dashed and dashed-dotted) lines for the inner and outer regions are shown in red and magenta, respectively.        }
    \label{fig:radio_xray_corr}
\end{figure*}

We examine the spatial relation between the radio and X-ray diffuse emission in RXJ1347. We extract the SB within the square boxes from the radio (uGMRT 700 MHz and MeerKAT 1.28 GHz) and X-ray (\textit{Chandra}) images. The box size is equal to the resolution ($9\arcsec$) of the radio images. The boxes in black and magenta are placed within the $3\sigma$ X-ray contour that fully covers the area of the radio emission (see Figs.~\ref{fig:Chandra_MeerKAT} and \ref{fig:radio_xray_corr}). The regions of discrete sources (i.e., radio galaxies and the SE excess emission; see Fig.~\ref{fig:radio_xray_corr}, left panel) are masked and are excluded from the analysis. In Fig.~\ref{fig:radio_xray_corr} (\textit{middle} panel), we present the scatter plot of the radio and X-ray SB extracted from the square boxes. The data points that have SB below $2\sigma$ in the radio maps are considered as non-detection and are indicated by the downward arrows in Fig.~\ref{fig:radio_xray_corr} (middle panel).

We fit Eq.~\ref{eq:radio_xray} with the extracted radio and X-ray SB within the boxes using the \texttt{linmix}\footnote{\url{https://linmix.readthedocs.io}} package that implements a hierarchical Bayesian model taking into account the uncertainties in both dependent and independent variables  \citep{Kelly2007}. In addition, \texttt{linmix} allows the fitting with data sets consisting of non-detections or missing data points. We obtained a sub-linear slope of $b_{\rm 700\,MHz}=0.82\pm0.04$ for the uGMRT 700 MHz -- X-ray data set and a linear slope of $b_{\rm 1.28\,GHz}=0.98\pm0.03$ for the MeerKAT 1.28 GHz -- X-ray data set. 

The radio -- X-ray SB scatter plot in Fig.~\ref{fig:radio_xray_corr} (middle panel)
shows possible different slopes in the inner and outer regions of the cluster. To examine this possibility, we separately fit the power-law relation in Eq.~\ref{eq:radio_xray} with the sub-datasets for the inner and outer clusters. The inner and outer regions are shown in Fig.~\ref{fig:radio_xray_corr} (left panel) in magenta and black, respectively. We obtain the best-fit slopes for the sub-datasets and show them in Table~\ref{tab:spatial_fitting_rx}. The best-fit slope of $b_{\rm 700\,MHz}=0.87\pm0.09$ for the inner region is steeper than that of $b_{\rm 700\,MHz}=0.56\pm0.06$ for the outer region. However, across the large area of the diffuse emission the slope at 1.28~GHz remains unchanged (i.e., $b_{\rm 1.28\,GHz}=0.91\pm0.09$ for the centre and $b_{\rm 1.28\,GHz}=0.89\pm0.05$ for the outskirts).

The correlation between spectral index and X-ray SB has been reported in a number of clusters \citep[e.g.][]{Botteon2020a,Rajpurohit2021,Biava2021}. The relation is described as follow,
\begin{equation}
\label{eq:alpha_xray}
\alpha = a' + b' \log {I_X},
\end{equation}
where $a'$ and $b'$ are free parameters. To examine the correlation in RXJ1347, we make a scatter plot between the 700~MHz--1.28~GHz spectral index and X-ray SB in Fig.~\ref{fig:radio_xray_corr} (right panel). The square regions where the data is extracted are identical to those used in the radio -- X-ray SB analysis (see Fig.~\ref{fig:radio_xray_corr}, left panel). Interestingly, the scatter plot shows two distinct regions (i.e., in red and magenta) with different slopes for the X-ray SB below and above $I_{X}=12.6\times10^{-6} \,{\rm cts \, s^{-1} \, arcmin^{-2} \, cm^{-2}}$ (or $\log{(I_{X})} = -5.9$). The data points with low X-ray SB (i.e., outer region of the cluster) follow a super-linear relation with a slope of $b'=1.22\pm0.23$, indicating a tight correlation between the spectral index and X-ray SB. However, the correlation slope flattens to $b'=0.16\pm0.14$ in the inner region (i.e., high X-ray SB), implying extremely weak $\alpha-I_X$ correlation.

\begin{table}
	\centering
	\caption{Best-fit slope for the spatial radio vs X-ray SB relation.}
	\begin{tabular}{llc}
		\hline\hline
		Relation & Region & slope \\ \hline
		uGMRT -- \textit{Chandra}    & All & $0.82\pm0.04$ \\
        & Inner & $0.87\pm0.09$ \\
        & Outer & $0.56\pm0.06$   \\  
        \hline
        MeerKAT -- \textit{Chandra}    & All & $0.98\pm0.03$ \\
        & Inner & $0.91\pm0.09$ \\
         & Outer &  $0.89\pm0.05$  \\     
        \hline    
        $\alpha_{\rm 700\,MHz}^{1.28\,GHz}$ -- \textit{Chandra} & Inner & $0.16\pm0.14$ \\
        & Outer & $1.22\pm0.23$\\
        \hline\hline
	\end{tabular}  
	\label{tab:spatial_fitting_rx}
\end{table}

\section{Discussion}
\label{sec:disc}

\subsection{Multi-component diffuse emission}
\label{sec:dis_multicomponent}

Mini-halos and radio halos are usually not observed in the same galaxy cluster. While mini-halos are pre-dominantly found in relaxed cool-core systems, radio halos are often associated with dynamically-disturbed galaxy clusters. Thanks to advancements in radio telescopes, particularly at sub-GHz frequencies, fainter diffuse steep-spectrum ($\alpha\lesssim-1$) radio emission has been recently observed in several systems. This has led to the detection of multiple central diffuse radio components within a handful of galaxy clusters. For instance, a combination of mini-halo and halo-like emission has been found in some galaxy clusters with LOFAR at 144~MHz \citep{Savini2019, Lusetti2023, Biava2024,Weeren2024}. These include the dynamically-relaxed clusters hosting cool cores (PSZ1~G139.61+24, RX~J1720.1+2638, PSZ1~G139.61+24, Abell~1068, MS~1455.0+2232, RX~J1720.1+2638, and Perseus cluster) and non-cool cores (Abell~1413). \cite{Cuciti2022a} also discovered mega- ($\gtrsim$2~Mpc) radio halos enveloping ($\lesssim$1~Mpc) radio halos in some merging galaxy clusters (Abell~665, Abell~697, Abell~2218, and ZwCl~0634.1+4750) using the LoTSS 144~MHz data. 

Our deep MeerKAT observations of RXJ1347 at 1.28~GHz show that the radial radio SB profile is consistent with a double exponential model (see Sec.~\ref{sec:res_sb_profiles}). The $e$-folding radii for the inner ($r_{e, {\rm inner}}=43\pm5$~kpc) and outer ($r_{e, {\rm outer}}=204\pm58$~kpc) components are within the typical ranges for those in mini-halos and halos, respectively. Previous observations found that the $e$-folding radius is typically smaller than $\sim$100~kpc for mini-halos and is larger, up to $\sim$500~kpc, for radio halos \citep[e.g.][]{Murgia2009,Lusetti2023, Biava2024,Balboni2024}. 
In addition, the best-fit SB peak for the inner component of RXJ1347 is significantly higher than that for the outer component, a factor of 20 at 1.28~GHz. In other similar cases hosting both a mini-halo and a halo, this ratio at 144~MHz ranges between 7 in MS~1455.0+2232 and 340 in RX~J1720.1+2638 \citep[e.g.][]{Lusetti2023,Biava2024}.
Moreover, the different dependence of the spectral index on the X-ray SB for the inner and outer regions through the $\alpha-I_{X}$ slopes ($b_{\rm inner}'=0.16\pm0.14$ and $b_{\rm outer}'=1.22\pm0.23$, respectively) indicates different nature of the diffuse components. 
Furthermore, the flux densities for the mini-halo and halo we obtained in Subsect.~\ref{sec:res_sb_profiles} are translated to volume emissivities of $(6.0\pm2.2)\times10^{-40} \, {\rm erg \, s^{-1}\,Hz^{-1}\,cm^{-3}}$ and $(6.7\pm6.8)\times10^{-42} \, {\rm erg \, s^{-1}\,Hz^{-1}\,cm^{-3}}$, respectively. These are in line with the values of $\sim10^{-40} {\rm erg \, s^{-1}\,Hz^{-1}\,cm^{-3}}$ and $\sim10^{-42} \, {\rm erg \, s^{-1}\,Hz^{-1}\,cm^{-3}}$ for mini-halos and halos reported in literature \citep[e.g.][]{Murgia2009}. 
In short, these results suggest that the cool-core galaxy cluster RXJ1347 hosts, both, a radio mini-halo and a  radio halo. 

\subsection{Particle acceleration mechanisms}
\label{sec:dis_mechanisms}

The spatial correlation between radio and X-ray emissions indicates a connection between the cluster's non-thermal (CRes and magnetic fields) and its thermal components. This correlation has proven to be a valuable tool for investigating the mechanisms driving the acceleration of relativistic electrons in the ICM \citep[e.g.][]{Govoni2001a}. In a merging galaxy cluster, a fraction of the gravitational energy dissipated during the collision can be converted into non-thermal components through the acceleration of CRes and the amplification of magnetic fields. The radio emissivity is given by,
\begin{equation}
    j_{R} \propto N_0 B^{\alpha+1} \nu^\alpha,
    \label{eq:emissivity_R}
\end{equation}
where $B$ is the magnetic field strength, $\nu$ is the frequency, and $\alpha=\nicefrac{(1-\delta)}{2}$ is the spectral index with $\delta$ being the index of the power-law energy distribution of  $N(\epsilon)d\epsilon=N_0\epsilon^{-\delta}d\epsilon$; here $\epsilon$ is the energy and $N_0$ is the normalisation \citep[e.g.][]{Govoni2004}. The emissivity of the thermal Bremsstrahlung producing X-rays is a function of the thermal electron density $n_e$ and temperature $k T_e$,
\begin{equation}
    j_{X} \propto n_e (k T_e)^{1/2}.
    \label{eq:emissivity_X}
\end{equation}
Hence, a connection between the radio and X-ray emissions may suggest a relationship between the non-thermal and thermal components of the cluster,
\begin{equation}
    N_0 B^{\alpha+1} \nu^\alpha \propto n_e (k T_e)^{\nicefrac{1}{2}}.
    \label{eq:RX_proto}
\end{equation}
It is important to note that the spatial distribution of the spectral index of the radio-emitting electrons also plays a significant role in this correlation, a factor that is often overlooked.

In the central mini-halo region of RXJ1347, we find a best-fit slope of $b_{\rm 700\,MHz}=0.87\pm0.09$ for the $I_R-I_X$ correlation. At the higher frequency of 1.28~GHz, the slope increases slightly to $b_{\rm 1.28\,GHz}=0.91\pm0.09$, although it remains within the $1\sigma$ uncertainty. While the high-frequency slope does not clearly indicate a sub-linear relation, the slope at the low frequency shows a sub-linear relation. To our knowledge, this mini-halo is the first to show a sub-linear slope in the radio and X-ray SB correlation. 

A recent study by \cite{Riseley2024} reports a sub-linear slope for the $I_R-I_X$ correlation for the central region of the merging galaxy cluster Abell 2142. However, the nature of the diffuse radio emission in this cluster remains uncertain; it could be a mix of a mini-halo and a halo \citep{Venturi2017} or simply a multi-component radio halo \citep{Bruno2023a}. Previous studies of confirmed mini-halos have consistently found linear or super-linear $I_R-I_X$ relations \citep{Biava2021,Ignesti2020,Riseley2022a,Lusetti2023}, which implies that the hadronic model for the acceleration of the CRes cannot be ruled out in these systems \citep{Govoni2001a}. 

The sub-linear relation observed in the RXJ1347 mini-halo suggests that the particle acceleration mechanism may involve more than just hadronic processes. It is possible that additional mechanisms, such as turbulence, also contribute to the diffuse radio emission. Thus, a combination of particle acceleration processes, including both hadronic and turbulent processes, might be at play in this mini-halo. 

In the outer halo region, we observe a slope of $b_{\rm 700\,MHz}=0.56\pm0.06$ at 700~MHz, which increases to $b_{\rm 1.28\,GHz}=0.89\pm0.05$ at the higher frequency. Both slopes are sub-linear, consistent with previous studies that have reported sub-linear correlations between the radio and X-ray SB emission in dynamically-disturbed galaxy clusters \citep[e.g.][]{Giacintucci2005,Cova2019,Xie2020a,Hoang2021b,Santra2024}. This sub-linear scaling in radio halos suggests that purely hadronic processes are unlikely to account for the observed correlation, as such processes would require an non-physically high contribution from CRps \citep[e.g.][]{Brunetti2014}.

An investigation by \cite{Balboni2024} finds that the non-thermal and thermal relation in five clusters hosting radio halos, using the LoTSS 144~MHz and CHEX-MATE (Cluster HEritage project with XMM-Newton - Mass Assembly and Thermodynamics at the Endpoint of structure formation; \citealt{Arnaud2021}) data. The authors find that the $I_R-I_X$ correlation is sub-linear and varies with radius, differing across clusters. In two relatively relaxed clusters, Abell~2244 and Abell~2409, the slope decreases in the outer region, indicating that the diffuse radio emission flattens with radius in these clusters. This pattern is also observed in RXJ1347 at 700~MHz, where the slope decreases from $0.87\pm0.09$ in the inner region to $0.56\pm0.06$ in the outer region. However, at higher frequency of 1.28~GHz the slope remains  unchanged with radius, around $0.9\pm0.1$.

The spectrum steepens from $\sim$$-1$ in the centre to $\sim$$-3$ in the outskirts of the cluster (see Fig.~\ref{fig:radio_xray_corr}, right panel). This could be due to the lower levels of turbulent energy flux that is channelled into the particle acceleration and is typically found in the outskirts of relaxed clusters. Reversely, as moving towards the cluster centre, turbulence is expected to increase, possibly resulting from past merging activities. Alternatively, the radial spectral steepening could be caused by the declination of magnetic field strength in the cluster outskirts \citep[e.g.][]{Brunetti2001,Bonafede2022}. Within the central mini-halo region, the spectral index shows a slight increase towards the centre, but still remains relatively flat, at $\sim$$-1$. This spectral feature may suggest a connection between the mini-halo and the gas sloshing of around the central radio galaxy and/or the radio-emitting CRes are additionally generated through the hadronic processes.

\section{Conclusions}
\label{sec:conc}

We provide a detailed analysis of multi-wavelength radio observations of the most X-ray luminous galaxy cluster RXJ1347, using the data from the MeerKAT at 1.28~GHz and the uGMRT at 700~MHz and 1.26~GHz. In combination with the \textit{Chandra} archival data, we investigate the correlation between the cluster's diffuse radio and X-ray diffuse emission. The key findings are summarized below.  

\begin{enumerate}

\item We have confirmed the existence of diffuse radio emission in the cluster's central region. Our observations reveal that the diffuse emission extends over much larger area, approximately 1~Mpc, compared to the previously reported extent of 640~kpc in literature.\\

\item The radial SB of the diffuse radio emission can be modelled by a double exponential model, with the inner component being notably brighter and smaller than the outer one. Based on their brightness, size, and emissivity, the inner component is classified as a mini-halo, while the outer component fits the profile of a radio halo.\\

\item Our radio and X-ray point-to-point analysis reveals distinct characteristics for the radio mini-halo and halo. For the mini-halo, we observe a sub-linear slope of $b_{\rm 700\,MHz}=0.87\pm0.09$ at 700~MHz; however, the slope at 1.28~GHz, $b_{\rm 1.28\,GHz}=0.91\pm0.09$, does not show a clear sub-linear trend. In contrast, the halo exhibits a sub-linear relation at both frequencies, with a flatter slope at 700~MHz ($b_{\rm 700\,MHz}=0.56\pm0.06$) compared to the steeper slope at 1.28~GHz ($b_{\rm 1.28~GHz}=0.89\pm0.05$). \\

\item The correlation between the spectral index and X-ray emission reveals distinct trends for the mini-halo and halo. In the mini-halo region, the spectral index shows a slight increase towards the cluster centre, with a slope of $b'_{\rm inner}=0.16\pm0.14$. In contrast, the halo has a clear spectral steepening, with a slope of $b'_{\rm outer}=1.22\pm0.23$. These findings indicate that different mechanisms are likely responsible for the formation of the mini-halo and halo within the cluster.\\

\item Our results on the sub-linear slopes observed in the radio versus X-ray relation for the radio halo are in line with the re-acceleration scenario in which the radio-emitting CRes are energized by turbulent processes. The observed spectral index steepens with radial distance, which we interpret as a result of turbulent energy from a past merger decreasing with distance from the cluster centre. 
In the mini-halo, however, the spectral index remains relatively constant with respect to the X-ray SB, indicating that the acceleration of the radio-emitting CRes is uniformly distributed across the mini-halo region, which spans approximately $\sim$100~kpc in radius. It is not entirely clear whether this acceleration is primarily due to hadronic processes or re-acceleration origin. The slightly sub-linear slope in the radio versus X-ray relation suggests that while the turbulent model might  contribute to the (re-)acceleration of the CRes in the mini-halo, it could be working in conjunction with hadronic processes, as has been proposed in other galaxy clusters.

\end{enumerate}

\begin{acknowledgements}
DNH is supported by the Deutsche Forschungsgemeinschaft (DFG, German Research Foundation) under research unit FOR 5195: "Relativistic Jets in Active Galaxies"). DNH, AB acknowledge support from the ERC through the grant ERC-Stg DRANOEL n. 714245.
MB acknowledges funding by the DFG under Germany's Excellence Strategy -- EXC 2121 "Quantum Universe" --  390833306.      
The MeerKAT telescope is operated by the South African Radio Astronomy Observatory, which is a facility of the National Research Foundation, an agency of the Department of Science and Innovation.
We thank the staff of the GMRT who made these observations possible. The GMRT is run by the National Centre for Radio Astrophysics of the Tata Institute of Fundamental Research.
The scientific results reported in this article are based  on data obtained from the \textit{Chandra} Data Archive.  This research has made use of software provided by the \textit{Chandra} X-ray Center (CXC) in the application packages CIAO, ChIPS, and Sherpa.
\end{acknowledgements}

\section*{ORCID IDs}
A. Bonafede \orcidlink{0000-0002-5068-4581} \href{https://orcid.org/0000-0002-5068-4581}{https://orcid.org/0000-0002-5068-4581} \\
M. Br\"uggen \orcidlink{0000-0002-3369-7735} \href{https://orcid.org/0000-0002-3369-7735}{https://orcid.org/0000-0002-3369-7735} \\
G. Brunetti \orcidlink{0000-0003-4195-8613 } \href{https://orcid.org/0000-0003-4195-8613 }{https://orcid.org/0000-0003-4195-8613 } \\
D. N. Hoang \orcidlink{0000-0002-8286-646X} \href{https://orcid.org/0000-0002-8286-646X}{https://orcid.org/0000-0002-8286-646X} \\
A. Liu \orcidlink{0000-0003-3501-0359} \href{https://orcid.org/0000-0003-3501-0359}{https://orcid.org/0000-0003-3501-0359} \\
E. Bulbul \orcidlink{0000-0002-7619-5399} \href{https://orcid.org/0000-0002-7619-5399}{https://orcid.org/0000-0002-7619-5399} \\
G. Di Gennaro \orcidlink{0000-0002-8648-8507} \href{https://orcid.org/0000-0002-8648-8507}{https://orcid.org/0000-0002-8648-8507} \\
P. Koch \orcidlink{0000-0003-2777-5861} \href{https://orcid.org/0000-0003-2777-5861}{https://orcid.org/0000-0003-2777-5861} \\
C. J. Riseley \orcidlink{0000-0002-3369-1085} \href{https://orcid.org/0000-0002-3369-1085}{https://orcid.org/0000-0002-3369-1085} \\
H. J. A. R\"ottgering \orcidlink{0000-0001-8887-2257} \href{https://orcid.org/0000-0001-8887-2257}{https://orcid.org/0000-0001-8887-2257} \\
R. J. van Weeren \orcidlink{0000-0002-0587-1660} \href{https://orcid.org/0000-0002-0587-1660}{https://orcid.org/0000-0002-0587-1660} \\    
%
\bibliographystyle{aa} 
\bibliography{RXJ1347} 
%

%
%
%
%

\begin{appendix} 
\section{Discrete source subtraction}
\label{sec:source_sub}
\begin{figure}[h]
\centering 
    \includegraphics[width=0.45\columnwidth]{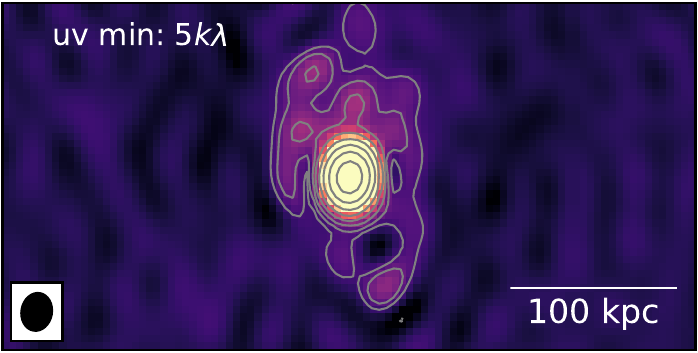}  
    \includegraphics[width=0.45\columnwidth]{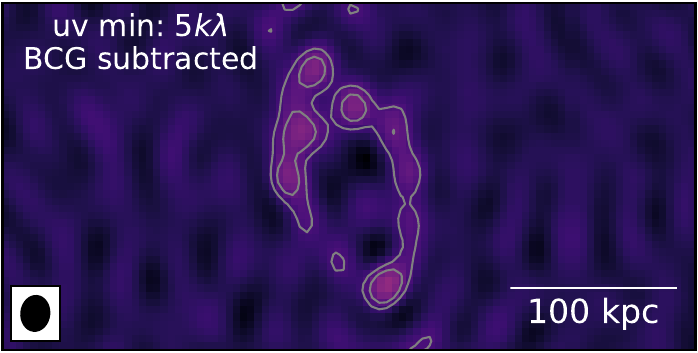} \\
    \includegraphics[width=0.45\columnwidth]{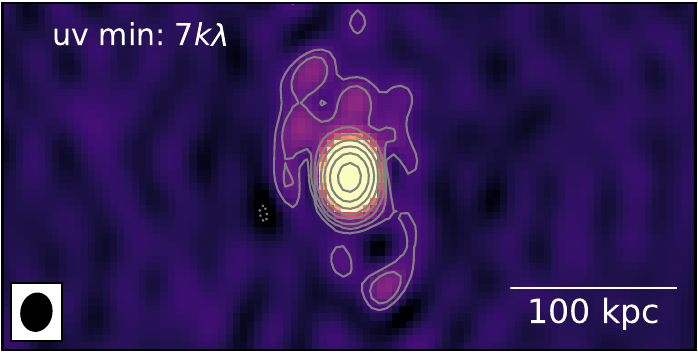}  
    \includegraphics[width=0.45\columnwidth]{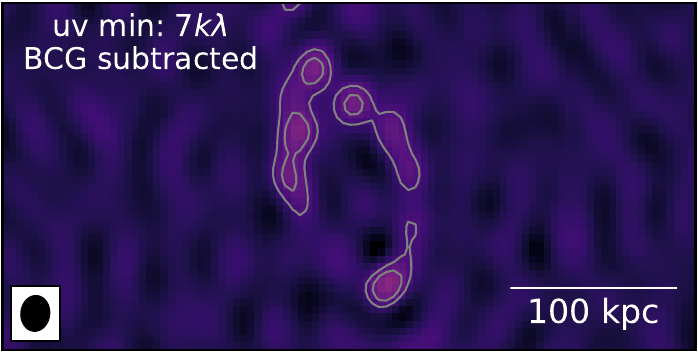} \\ 
    \includegraphics[width=0.45\columnwidth]{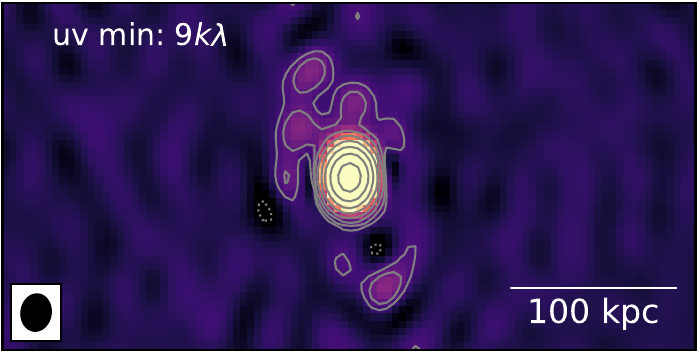}  
    \includegraphics[width=0.45\columnwidth]{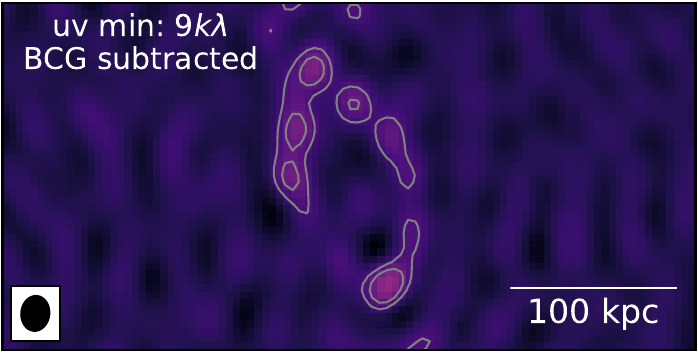} \\
    \includegraphics[width=0.45\columnwidth]{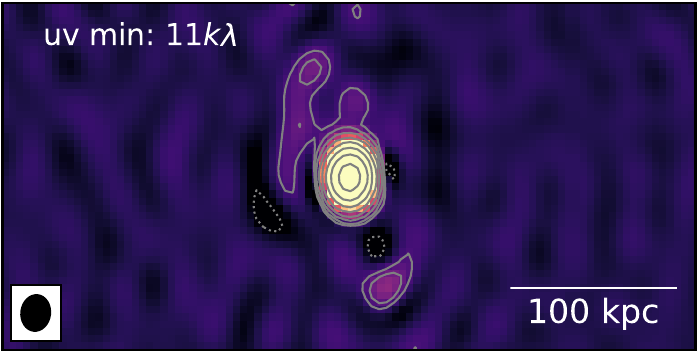}  
    \includegraphics[width=0.45\columnwidth]{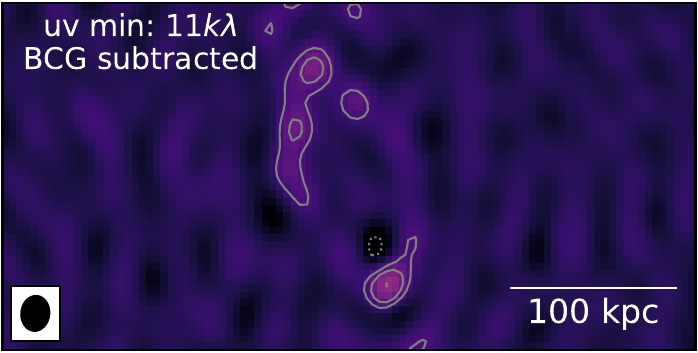} \\
    \includegraphics[width=0.45\columnwidth]{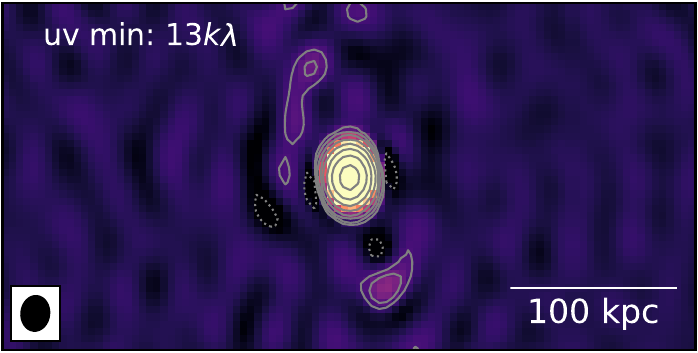}  
    \includegraphics[width=0.45\columnwidth]{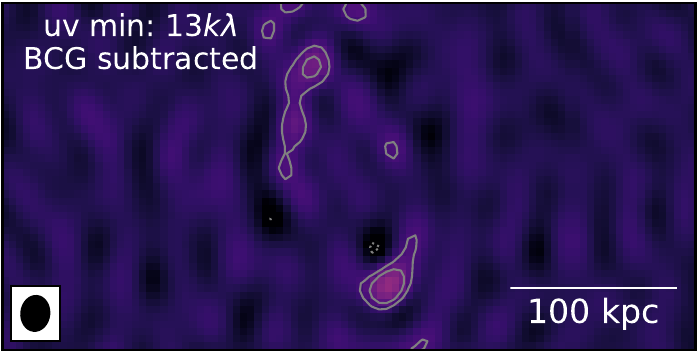} \\
    \includegraphics[width=0.45\columnwidth]{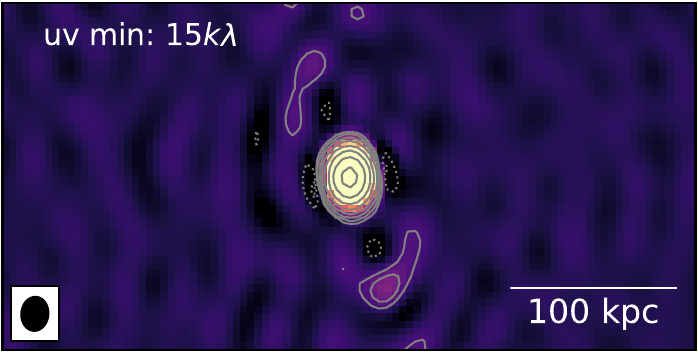}  
    \includegraphics[width=0.45\columnwidth]{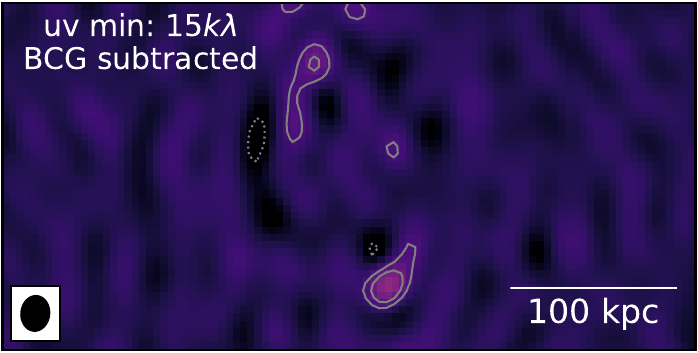} \\
    \includegraphics[width=0.45\columnwidth]{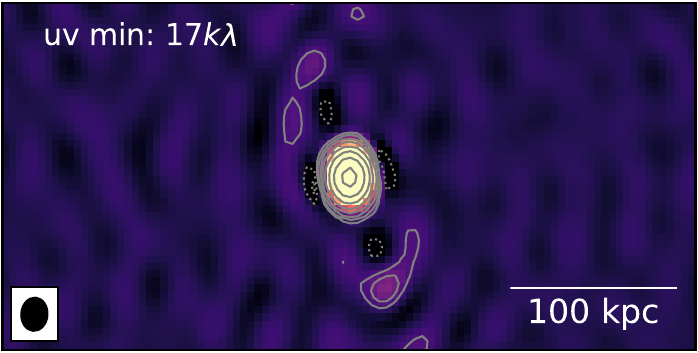}  
    \includegraphics[width=0.45\columnwidth]{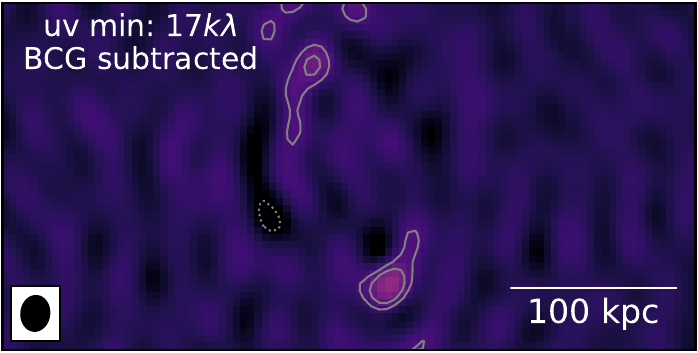} \\
    \includegraphics[width=0.45\columnwidth]{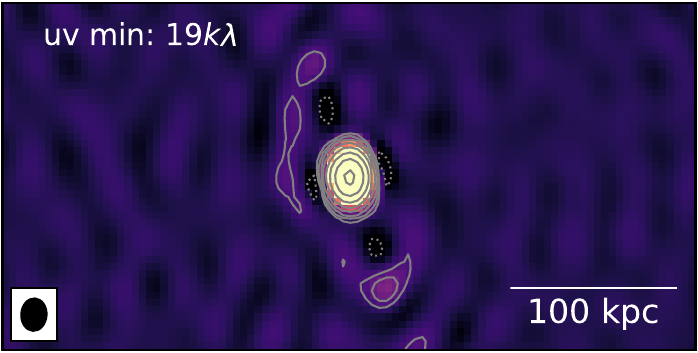}  
    \includegraphics[width=0.45\columnwidth]{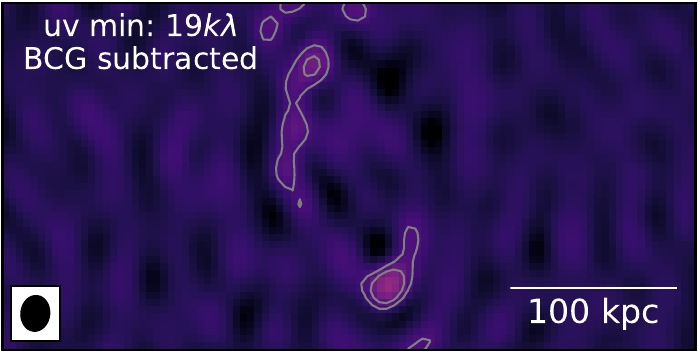} \\
    \caption{MeerKAT images of the centre region of RXJ1347 were created using a range of uv minimums from 5~$k\lambda$ to 19~$k\lambda$. The first row shows the detection of the central BCG (A), which is removed in the second row.}
    \label{fig:BCG_sub_uv_mins}
\end{figure}

\begin{figure}
\centering 
    \includegraphics[width=0.8\columnwidth]{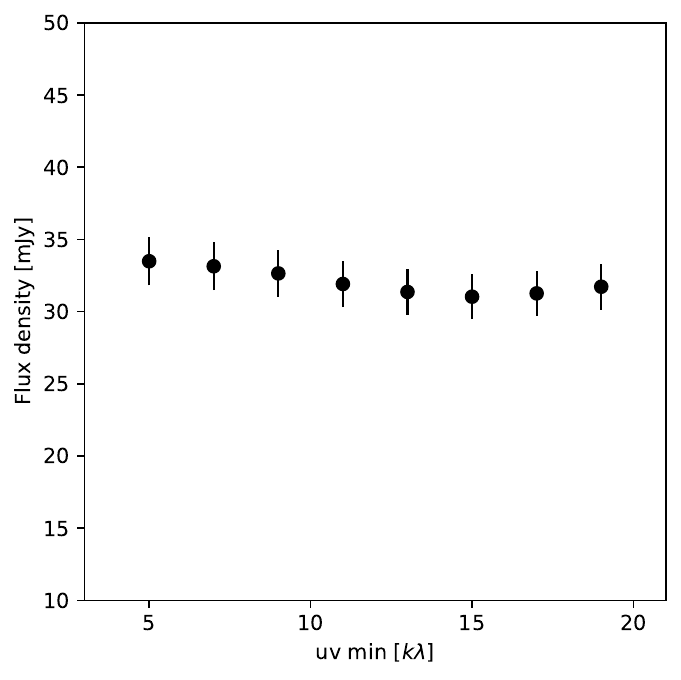}
    \caption{The flux density of the central BCG (A) measured from the images created with various $uv$ minimums (see Fig.~\ref{fig:BCG_sub_uv_mins}).}
    \label{fig:BCG_flux_uvmins}
\end{figure}

Figs.~\ref{fig:BCG_sub_uv_mins} (left panels) show the radio images of the cluster RXJ1347 which  are obtained with different $uv$ minimum cut-off between $5\,{\rm k\lambda}$ and $19\,{\rm k\lambda}$. The image with $uv$ cut-off of $13\,{\rm k\lambda}$ shows clear separation between the central BCG and the surrounding diffuse emission. The BCG is subtracted from the data and are shown in the right  panels. In Fig.~\ref{fig:BCG_flux_uvmins} we plot the radio flux density of the BCG measured measured from these $uv$ cut-off images. 

\begin{figure}
\centering 
    \includegraphics[width=0.4\columnwidth]{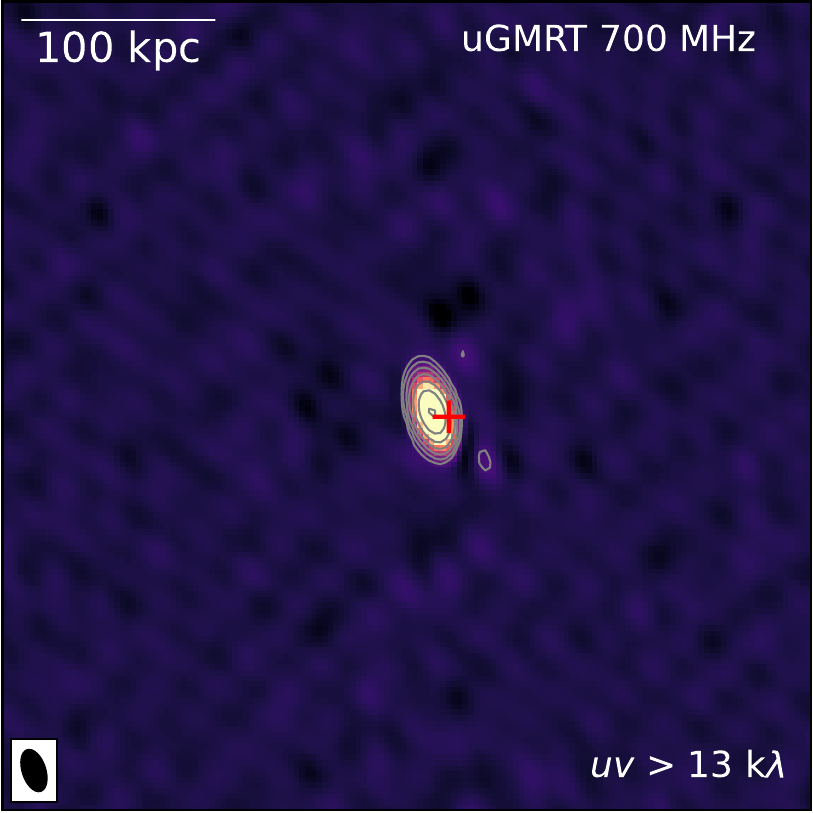}  \hfil 
    \includegraphics[width=0.4\columnwidth]{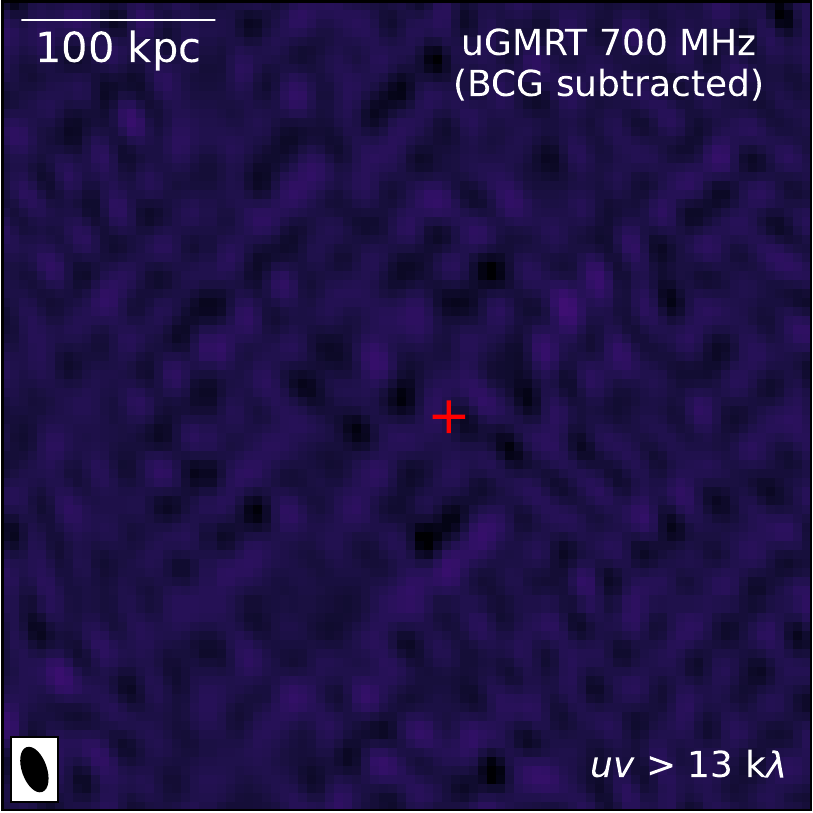}  \\ 
    \includegraphics[width=0.4\columnwidth]{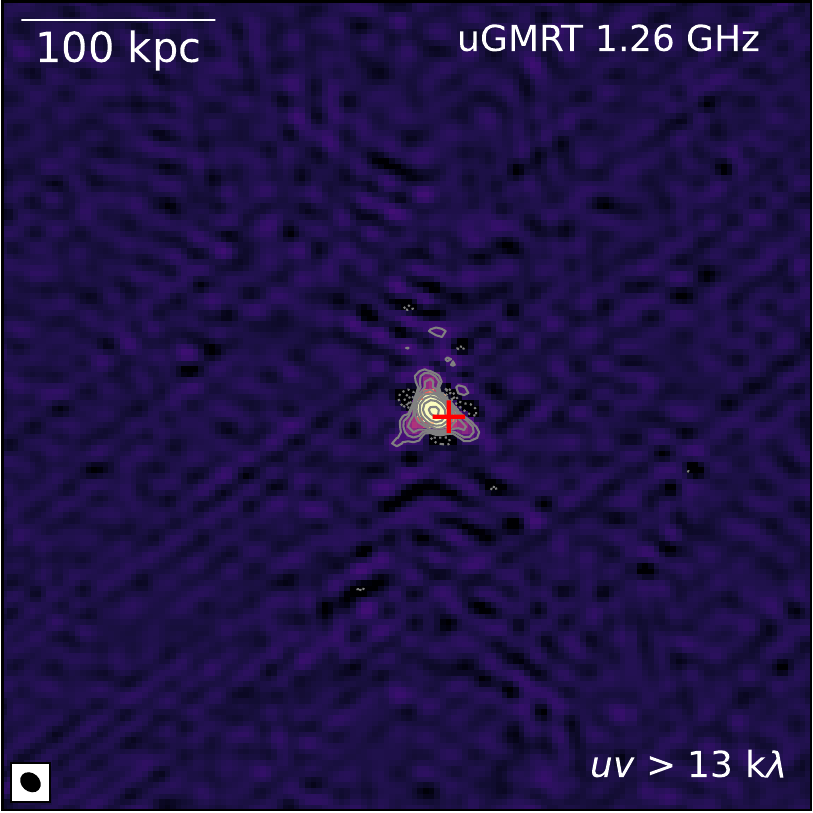} \hfil
    \includegraphics[width=0.4\columnwidth]{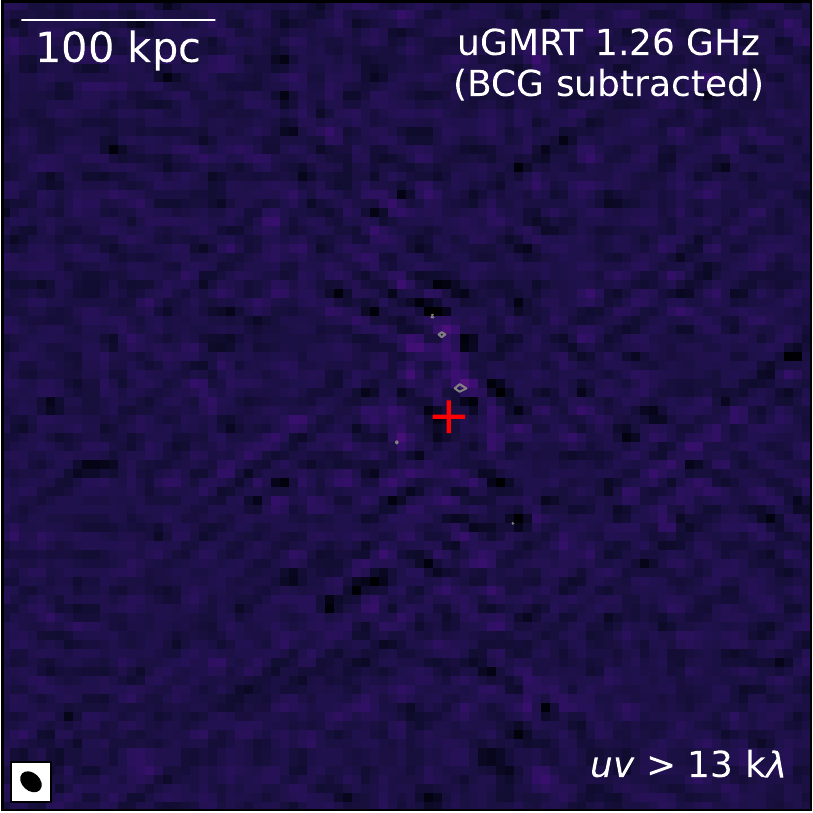} \\
    \includegraphics[width=0.4\columnwidth]{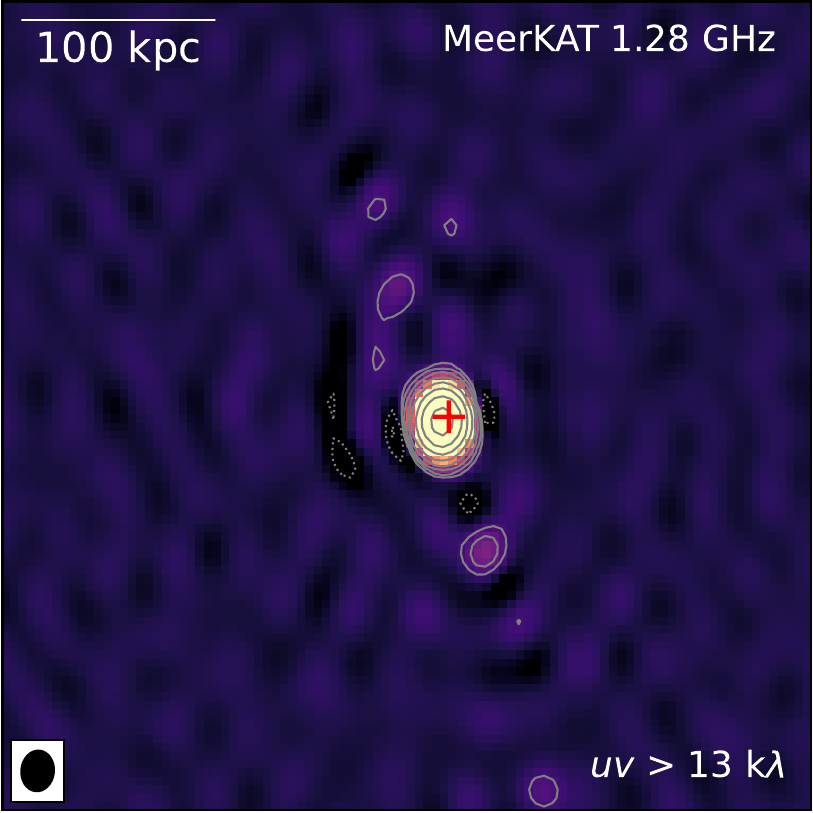} \hfil
    \includegraphics[width=0.4\columnwidth]{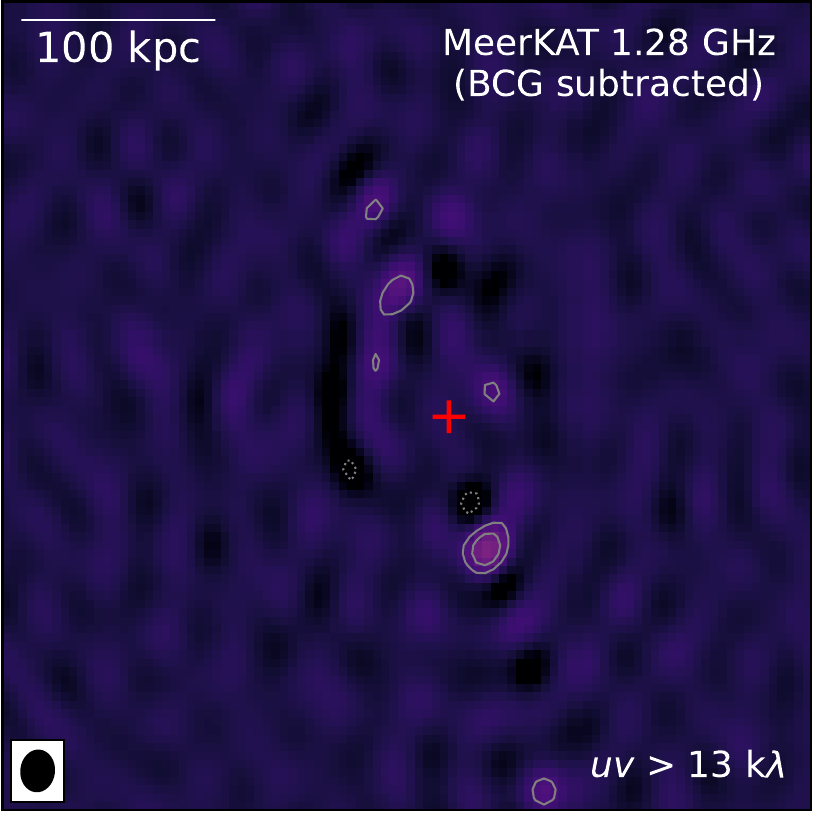} \\
    \caption{High-resolution images of the cluster central region with (left column) and without (right column) the radio galaxy. The synthesis beams are shown in the bottom-left corners. The X-ray peak location is marked with the red cross. The contours start at $5\sigma$ and are spaced by a factor of 2.}
    \label{fig:BCG_sub_uGMRT_MeerKAT}
\end{figure}

The models of the central radio galaxy that is obtained with $uv$ minimum cut-off of $13\,{\rm k\lambda}$ are subtracted from the MeerKAT and uGMRT $uv$ data. In Fig.~\ref{fig:BCG_sub_uGMRT_MeerKAT}, we show the high-resolution images of the cluster central region with (left panels) and without (right panels) the BCG.

\end{appendix}
  	
\end{document}